\documentclass[apj,iop]{emulateapj}
\usepackage{natbib,graphicx,amsmath}
\pdfoutput=1

\newcommand{\Om}{$\Omega_\textrm m$}
\newcommand{\Asz}{$A_\textrm{SZ}$}
\newcommand{\Bsz}{$B_\textrm{SZ}$}
\newcommand{\Csz}{$C_\textrm{SZ}$}
\newcommand{\Dsz}{$D_\textrm{SZ}$}
\newcommand{\Adisp}{$A_{\sigma_v}$}
\newcommand{\sigmav}{$\sigma_v$}
\newcommand{\Yx}{$Y_\textrm X$}
\newcommand{\widthparam}{$\sigma_8(\Omega_\textrm m/0.27)^{0.3}$}

\newcommand{\planck}{\textit{Planck}}
\newcommand{\chandra}{\textit{Chandra}}
\newcommand{\xmm}{\textit{XMM-Newton}}
\newcommand{\spitzer}{\textit{Spitzer}}

\newcommand{\SPTcl}{SPT$_\textrm{CL}$}
\newcommand{\Nxiz}{$N(\xi,z)$}
\newcommand{\LCDM}{$\Lambda$CDM}
\newcommand{\summnu}{$\sum m_\nu$}
\newcommand{\vect}{\boldsymbol}

\def\Munich{1}
\def\ExcellenceCluster{2}
\def\MPE{3}
\def\UChicago{4}
\def\CfA{5}
\def\MIT{6}
\def\Harvard{7}
\def\FNAL{8}
\def\KICPChicago{9}
\def\AAUChicago{10}
\def\PhysicsUChicago{11}
\def\ANL{12}
\def\Miss{13}
\def\EFIChicago{14}
\def\NIST{15}
\def\PUC{16}
\def\Caltech{17}
\def\McGill{18}
\def\illast{19}
\def\illphy{20}
\def\Berkeley{21}
\def\UFlorida{22}
\def\Colorado{23}
\def\KIPAC{24}
\def\Stanford{25}
\def\Davis{26}
\def\LBNL{27}
\def\Arizona{28}
\def\Michigan{29}
\def\Minnesota{30}
\def\UMelbourne{31}
\def\STScI{32}
\def\CaseWestern{33}
\def\SAIC{34}
\def\LLNL{35}
\def\Dunlap{36}
\def\Toronto{37}
\def\BCCP{38}
\def\CTIO{39}

\altaffiltext{\Munich}{Department of Physics, Ludwig-Maximilians-Universit\"{a}t, Scheinerstr.\ 1, 81679 M\"{u}nchen, Germany}
\altaffiltext{\ExcellenceCluster}{Excellence Cluster Universe, Boltzmannstr.\ 2, 85748 Garching, Germany}
\altaffiltext{\MPE}{Max-Planck-Institut f\"{u}r extraterrestrische Physik, Giessenbachstr. 1,\ 85748 Garching, Germany}
\altaffiltext{\UChicago}{University of Chicago, 5640 South Ellis Avenue, Chicago, IL 60637, USA}
\altaffiltext{\CfA}{Harvard-Smithsonian Center for Astrophysics, 60 Garden Street, Cambridge, MA 02138, USA}
\altaffiltext{\MIT}{Kavli Institute for Astrophysics and Space Research, Massachusetts Institute of Technology, 77 Massachusetts Avenue, Cambridge, MA 02139,  USA}
\altaffiltext{\Harvard}{Department of Physics, Harvard University, 17 Oxford Street, Cambridge, MA 02138, USA}
\altaffiltext{\FNAL}{Fermi National Accelerator Laboratory, Batavia, IL 60510-0500, USA}
\altaffiltext{\KICPChicago}{Kavli Institute for Cosmological Physics, University of Chicago, 5640 South Ellis Avenue, Chicago, IL 60637, USA}
\altaffiltext{\AAUChicago}{Department of Astronomy and Astrophysics, University of Chicago, 5640 South Ellis Avenue, Chicago, IL 60637, USA}
\altaffiltext{\PhysicsUChicago}{Department of Physics, University of Chicago, 5640 South Ellis Avenue, Chicago, IL 60637, USA}
\altaffiltext{\ANL}{Argonne National Laboratory, High-Energy Physics Division, 9700 S. Cass Avenue, Argonne, IL, USA 60439, USA}
\altaffiltext{\Miss}{Department of Physics and Astronomy, University of Missouri, 5110 Rockhill Road, Kansas City, MO 64110, USA}
\altaffiltext{\EFIChicago}{Enrico Fermi Institute, University of Chicago, 5640 South Ellis Avenue, Chicago, IL 60637, USA}
\altaffiltext{\NIST}{NIST Quantum Devices Group, 325 Broadway Mailcode 817.03, Boulder, CO, USA 80305, USA}
\altaffiltext{\PUC}{Departamento de Astronomia y Astrosifica, Pontificia Universidad Catolica, Chile}
\altaffiltext{\Caltech}{California Institute of Technology, 1200 E. California Blvd., Pasadena, CA 91125, USA}
\altaffiltext{\McGill}{Department of Physics, McGill University, 3600 Rue University, Montreal, Quebec H3A 2T8, Canada}
\altaffiltext{\illast}{Astronomy Department, University of Illinois at Urbana-Champaign, 1002 W.\ Green Street, Urbana, IL 61801, USA}
\altaffiltext{\illphy}{Department of Physics, University of Illinois Urbana-Champaign, 1110 W.\ Green Street, Urbana, IL 61801, USA}
\altaffiltext{\Berkeley}{Department of Physics, University of California, Berkeley, CA 94720, USA}
\altaffiltext{\UFlorida}{Department of Astronomy, University of Florida, Gainesville, FL 32611, USA}
\altaffiltext{\Colorado}{Department of Astrophysical and Planetary Sciences and Department of Physics, University of Colorado, Boulder, CO 80309, USA}
\altaffiltext{\KIPAC}{Kavli Institute for Particle Astrophysics and Cosmology, Stanford University, 452 Lomita Mall, Stanford, CA 94305, USA}
\altaffiltext{\Stanford}{Department of Physics, Stanford University, 382 Via Pueblo Mall, Stanford, CA 94305, USA}
\altaffiltext{\Davis}{Department of Physics, University of California, One Shields Avenue, Davis, CA 95616, USA}
\altaffiltext{\LBNL}{Physics Division, Lawrence Berkeley National Laboratory, Berkeley, CA 94720, USA}
\altaffiltext{\Arizona}{Steward Observatory, University of Arizona, 933 North Cherry Avenue, Tucson, AZ 85721, USA}
\altaffiltext{\Michigan}{Department of Physics, University of Michigan, 450 Church Street, Ann Arbor, MI 48109, USA}
\altaffiltext{\Minnesota}{Physics Department, University of Minnesota, 116 Church Street S.E., Minneapolis, MN 55455, USA}
\altaffiltext{\UMelbourne}{School of Physics, University of Melbourne, Parkville, VIC 3010, Australia}
\altaffiltext{\STScI}{Space Telescope Science Institute, 3700 San Martin Dr., Baltimore, MD 21218, USA}
\altaffiltext{\CaseWestern}{Physics Department, Center for Education and Research in Cosmology and Astrophysics, Case Western Reserve University, Cleveland, OH 44106, USA}
\altaffiltext{\SAIC}{Liberal Arts Department, School of the Art Institute of Chicago, 112 S Michigan Ave, Chicago, IL 60603, USA}
\altaffiltext{\LLNL}{Institute of Geophysics and Planetary Physics, Lawrence, Livermore National Laboratory, Livermore, CA 94551, USA}
\altaffiltext{\Dunlap}{Dunlap Institute for Astronomy \& Astrophysics, University of Toronto, 50 St George St, Toronto, ON, M5S 3H4, Canada}
\altaffiltext{\Toronto}{Department of Astronomy \& Astrophysics, University of Toronto, 50 St George St, Toronto, ON, M5S 3H4, Canada}
\altaffiltext{\BCCP}{Berkeley Center for Cosmological Physics, Department of Physics, University of California, and Lawrence Berkeley National Labs, Berkeley, A 94720, USA}
\altaffiltext{\CTIO}{Cerro Tololo Inter-American Observatory, Casilla 603, La Serena, Chile}

\shorttitle{Mass Calibration and Cosmological Analysis of SPT-SZ Galaxy Clusters}
\shortauthors{Bocquet et al.}

\begin{document}

\title{Mass calibration and cosmological analysis of the SPT-SZ galaxy cluster \\
sample using velocity dispersion \sigmav\ and X-ray \Yx\ measurements}

\author{
S.~Bocquet\altaffilmark{\Munich,\ExcellenceCluster},
A.~Saro\altaffilmark{\Munich},
J.~J.~Mohr\altaffilmark{\Munich,\ExcellenceCluster,\MPE},
K.~A.~Aird\altaffilmark{\UChicago},
M.~L.~N.~Ashby\altaffilmark{\CfA},
M.~Bautz\altaffilmark{\MIT},
M.~Bayliss\altaffilmark{\CfA,\Harvard},
G.~Bazin\altaffilmark{\Munich},
B.~A.~Benson\altaffilmark{\FNAL,\KICPChicago,\AAUChicago},
L.~E.~Bleem\altaffilmark{\KICPChicago,\PhysicsUChicago,\ANL},
M.~Brodwin\altaffilmark{\Miss},
J.~E.~Carlstrom\altaffilmark{\KICPChicago,\AAUChicago,\PhysicsUChicago,\ANL,\EFIChicago}, 
C.~L.~Chang\altaffilmark{\KICPChicago,\ANL,\EFIChicago}, 
I.~Chiu\altaffilmark{\Munich},
H.~M.~Cho\altaffilmark{\NIST}, 
A.~Clocchiatti\altaffilmark{\PUC},
T.~M.~Crawford\altaffilmark{\KICPChicago,\AAUChicago},
A.~T.~Crites\altaffilmark{\KICPChicago,\AAUChicago,\Caltech},
S.~Desai\altaffilmark{\Munich,\ExcellenceCluster},
T.~de~Haan\altaffilmark{\McGill},
J.~P.~Dietrich\altaffilmark{\Munich,\ExcellenceCluster},
M.~A.~Dobbs\altaffilmark{\McGill},
R.~J.~Foley\altaffilmark{\CfA,\illast,\illphy},
W.~R.~Forman\altaffilmark{\CfA},
D.~Gangkofner\altaffilmark{\Munich,\ExcellenceCluster},
E.~M.~George\altaffilmark{\Berkeley},
M.~D.~Gladders\altaffilmark{\KICPChicago,\AAUChicago},
A.~H.~Gonzalez\altaffilmark{\UFlorida},
N.~W.~Halverson\altaffilmark{\Colorado},
C.~Hennig\altaffilmark{\Munich},
J.~Hlavacek-Larrondo\altaffilmark{\KIPAC,\Stanford},
G.~P.~Holder\altaffilmark{\McGill},
W.~L.~Holzapfel\altaffilmark{\Berkeley},
J.~D.~Hrubes\altaffilmark{\UChicago},
C.~Jones\altaffilmark{\CfA},
R.~Keisler\altaffilmark{\KICPChicago,\PhysicsUChicago},
L.~Knox\altaffilmark{\Davis},
A.~T.~Lee\altaffilmark{\Berkeley,\LBNL},
E.~M.~Leitch\altaffilmark{\KICPChicago,\AAUChicago},
J.~Liu\altaffilmark{\Munich,\ExcellenceCluster},
M.~Lueker\altaffilmark{\Caltech,\Berkeley},
D.~Luong-Van\altaffilmark{\UChicago},
D.~P.~Marrone\altaffilmark{\Arizona},
M.~McDonald\altaffilmark{\MIT},
J.~J.~McMahon\altaffilmark{\Michigan},
S.~S.~Meyer\altaffilmark{\KICPChicago,\AAUChicago,\PhysicsUChicago,\EFIChicago},
L.~Mocanu\altaffilmark{\KICPChicago,\AAUChicago},
S.~S.~Murray\altaffilmark{\CfA},
S.~Padin\altaffilmark{\KICPChicago,\AAUChicago,\Caltech},
C.~Pryke\altaffilmark{\Minnesota}, 
C.~L.~Reichardt\altaffilmark{\Berkeley,\UMelbourne}, 
A.~Rest\altaffilmark{\STScI}, 
J.~Ruel\altaffilmark{\Harvard}, 
J.~E.~Ruhl\altaffilmark{\CaseWestern}, 
B.~R.~Saliwanchik\altaffilmark{\CaseWestern}, 
J.~T.~Sayre\altaffilmark{\CaseWestern}, 
K.~K.~Schaffer\altaffilmark{\KICPChicago,\EFIChicago,\SAIC}, 
E.~Shirokoff\altaffilmark{\Berkeley,\Caltech}, 
H.~G.~Spieler\altaffilmark{\LBNL},
B.~Stalder\altaffilmark{\CfA},
S.~A.~Stanford\altaffilmark{\Davis,\LLNL},
Z.~Staniszewski\altaffilmark{\Caltech,\CaseWestern},
A.~A.~Stark\altaffilmark{\CfA}, 
K.~Story\altaffilmark{\KICPChicago,\PhysicsUChicago},
C.~W.~Stubbs\altaffilmark{\CfA,\Harvard}, 
K.~Vanderlinde\altaffilmark{\Dunlap,\Toronto},
J.~D.~Vieira\altaffilmark{\illast,\illphy},
A. Vikhlinin\altaffilmark{\CfA},
R.~Williamson\altaffilmark{\KICPChicago,\AAUChicago,\Caltech}, 
O.~Zahn\altaffilmark{\Berkeley,\BCCP}, and
A.~Zenteno\altaffilmark{\Munich,\CTIO}
}

\email{bocquet@usm.lmu.de}

\begin{abstract}

We present a velocity dispersion-based mass calibration of the South Pole Telescope Sunyaev-Zel'dovich effect survey (SPT-SZ) galaxy cluster sample. Using a homogeneously selected sample of 100 cluster candidates from 720~deg$^2$ of the survey along with 63 velocity dispersion ($\sigma_v$) and 16 X-ray $Y_\textrm{X}$ measurements of sample clusters, we simultaneously calibrate the mass-observable relation and constrain cosmological parameters. Our method accounts for cluster selection, cosmological sensitivity, and uncertainties in the mass calibrators. The calibrations using $\sigma_v$ and $Y_\textrm{X}$ are consistent at the $0.6\sigma$ level, with the $\sigma_v$ calibration preferring $\sim$16\% higher masses. We use the full SPT$_\textrm{CL}$ dataset (SZ clusters+$\sigma_v$+$Y_\textrm{X}$) to measure $\sigma_8(\Omega_\textrm m/0.27)^{0.3}=0.809\pm0.036$ within a flat $\Lambda$CDM model. The SPT cluster abundance is lower than preferred by either the WMAP9 or \textit{Planck}+WMAP9 polarization (WP) data, but assuming the sum of the neutrino masses is $\sum m_\nu=0.06$~eV, we find the datasets to be consistent at the 1.0$\sigma$ level for WMAP9 and 1.5$\sigma$ for \textit{Planck}+WP.  Allowing for larger $\sum m_\nu$ further reconciles the results.  When we combine the SPT$_\textrm{CL}$ and \textit{Planck}+WP datasets with information from baryon acoustic oscillations and supernovae Ia, the preferred cluster masses are $1.9\sigma$ higher than the $Y_\textrm{X}$ calibration and $0.8\sigma$ higher than the $\sigma_v$ calibration. Given the scale of these shifts ($\sim$44\% and $\sim$23\% in mass, respectively), we execute a goodness of fit test; it reveals no tension, indicating that the best-fit model provides an adequate description of the data. Using the multi-probe dataset, we measure $\Omega_\textrm m=0.299\pm0.009$ and $\sigma_8=0.829\pm0.011$.  Within a $\nu$CDM model we find $\sum m_\nu = 0.148\pm0.081$~eV. We present a consistency test of the cosmic growth rate using SPT clusters. Allowing both the growth index $\gamma$ and the dark energy equation of state parameter $w$ to vary, we find $\gamma=0.73\pm0.28$ and $w=-1.007\pm0.065$, demonstrating that the expansion and the growth histories are consistent with a $\Lambda$CDM Universe ($\gamma=0.55; \,w=-1$).

\end{abstract}



\section{Introduction}

Galaxy cluster surveys provide important insights into cosmological questions such as the nature of cosmic acceleration \citep{wang98,haiman01,holder01b,battye03,molnar04,wang04,lima07}, the Gaussian character of underlying density perturbations \citep{dalal08,cayon11,williamson11} and the cosmic growth rate \citep{rapetti13}. Because their distribution in mass and redshift depends on both the geometry of the Universe and the growth rate of structure, galaxy clusters are complementary to distance-based probes such as Type~Ia Supernovae \citep[e.g.,][]{sullivan11} and Baryon Acoustic Oscillations \citep[e.g.,][]{percival10}. Indeed, recent studies demonstrate the constraining power of galaxy clusters using real cluster samples in X-ray \citep[e.g.,][]{vikhlinin09b,mantz10a}, optical \citep[e.g.,][]{rozo10} and Sunyaev-Zel'dovich effect \citep[SZE; e.g.,][]{vanderlinde10,sehgal11,benson13,reichardt13,hasselfield13,planck13-20} surveys.

Today, the largest available cluster catalogs come from X-ray and optical surveys.  However, galaxy clusters can also be detected through their thermal SZE signature, which arises from the interaction of the cosmic microwave background (CMB) photons with the hot, ionized intracluster medium \citep{sunyaev72}. The surface brightness of the SZE signature is independent of redshift, and the integrated signature is expected to be a low-scatter mass proxy \citep{barbosa96,holder01b,motl05,nagai07,stanek10}.
Therefore, SZE cluster surveys with sufficient angular resolution are expected to generate nearly mass-limited samples extending to the highest redshifts at which clusters exist. Dedicated millimeter-wave SZE surveys over large areas of the sky are being carried out by the South Pole Telescope \citep[SPT,][]{carlstrom11}, the Atacama Cosmology Telescope \citep{fowler07}, and \planck\ \citep{planck11-13}.

The first cosmological analysis of an SPT cluster sample used 21 clusters selected from 178~deg$^2$ of survey data \citep{vanderlinde10}.
The observed SPT signal-to-noise $\xi$ was used as a proxy for cluster mass, assuming a relationship that was calibrated from simulations.
Using the same cluster sample, \cite{benson13} repeated the cosmological analysis using additional mass calibration from the X-ray observable $Y_\textrm X\equiv M_\textrm g T_\textrm X$, where $M_\textrm g$ is the intracluster gas mass and $T_\textrm X$ is the X-ray temperature.
The X-ray data were obtained for a sub-sample of 14 clusters using \chandra\ and \xmm\ \citep{andersson11}.
The combination of the cluster abundance measurements with CMB anisotropy data improved constraints on \Om\ and $\sigma_8$ by a factor of 1.5 over the results from CMB data alone \citep[WMAP7,][]{komatsu11}.
Most recently, \cite{reichardt13} analyzed a sample of 100 cluster candidates extracted from the first 720~deg$^2$ of the SPT-SZ survey, including X-ray data on the same 14 clusters. The uncertainty in the derived cosmological constraints was dominated by the systematic uncertainties in the mass calibration of the sample.

Given the importance of the cluster mass calibration, the SPT collaboration has undertaken a comprehensive follow-up program to make use of multiple mass measurement techniques to better characterize the SPT mass-observable relation. Our strategy is to obtain direct mass constraints from X-ray observations and cluster velocity dispersions, and these will be supplemented with mass constraints from weak lensing in future studies. Both velocity dispersions and weak lensing exhibit significant uncertainties on individual cluster mass measurements but can be studied in detail using $N$-body studies of structure formation in order to characterize and correct for the systematic biases \citep[e.g.,][]{white10,becker11,saro13}. Therefore, large ensembles of these measurements can be combined to deliver precise and accurate mass information. In a complementary fashion, the X-ray mass proxy \Yx\ is tightly correlated with the cluster virial mass, and can be calibrated using weak lensing or velocity dispersions to provide accurate and reasonably precise single cluster mass measurements \citep[e.g.,][]{sun09,vikhlinin09a,mantz10b}. In addition, we expect the small scatter X-ray observable to play an important role as we want to constrain not only the masses of our SPT clusters, but also the scatter about the SPT mass-observable relation. The latter plays a central role in the SPT cluster survey selection, and is critically important for the cosmological interpretation of the sample \citep[e.g.,][]{lima05}.

In this work, we report a detailed analysis of the SZE mass-observable relation calibration using the cluster sample of the 720~deg$^2$ SPT-SZ survey together with a subset of 64 SZE detected galaxy clusters with additional spectroscopic and/or X-ray observations. The cluster sample with its mass calibration data and external cosmological datasets are described in Section~\ref{sec:data}. In Section~\ref{sec:velocitydispersion} we summarize how velocity dispersions are used as mass calibrators, and largely follow the recent theoretical exploration of this issue \citep{saro13}. We present our analysis method in Section~\ref{sec:method}, and show how we tested it on simulated data. In Section~\ref{sec:results} we compare the X-ray and velocity dispersion constraints.  Because they are in good agreement, we combine them and present our best current constraints from SPT clusters alone assuming a flat \LCDM\ model, showing that these results are in agreement with constraints from external datasets. We then carry out a joint cosmological analysis that combines our SPT clusters with external data to deliver the tightest constraints on cluster masses and cosmological parameters. We also explore constraints on the sum of the neutrino masses, cosmic growth, and the Dark Energy equation of state parameter $w$. We review our conclusions in Section~\ref{sec:conclusions}.  

In this work, unless otherwise specified, we assume a flat \LCDM\ cosmology with massless neutrinos. Cluster masses refer to $M_{500,\textrm c}$, the mass enclosed within a sphere of radius $r_{500}$, in which the mean matter density is equal to 500 times the critical density. The critical density at the cluster's redshift is $\rho_\textrm{crit}(z) = 3H^2(z)/8\pi G$, where $H(z)$ is the Hubble parameter.


\section{Observations and Data}
\label{sec:data}

\subsection{South Pole Telescope Observations, Cluster Catalog, and Scaling Relations}
\label{sec:SZdata}
The SPT is a 10~m telescope located within 1~km of the geographical South Pole. From 2007 to 2011, the telescope was configured to observe in three millimeter-wave bands (centered at 95, 150, and 220~GHz). The majority of this period was spent on a survey of a contiguous 2500 deg$^2$ area within the boundaries 20h~$\leq$~R.A.~$\leq$~7h and $-65^\circ\leq\textrm{Dec.}\leq-40^\circ$, which we term the SPT-SZ survey. The survey was completed in November 2011, and achieved a fiducial depth of 18~$\mu$K-arcmin in the 150~GHz band.
Details of the survey strategy and data processing can be found in \cite{schaffer11}.

Galaxy clusters are detected via their thermal SZE signature in the 95 and 150~GHz maps. These maps are created using time-ordered data processing and map-making procedures equivalent to those described in \cite{vanderlinde10}, and clusters are extracted from the multi-band data as in \cite{williamson11, reichardt13}. A multi-scale matched-filter approach is used for cluster detection \citep{melin06}. The observable of the cluster SZE signal is $\xi$, the detection significance maximized over all filter scales. Because of the impact of noise biases, a direct scaling relation between $\xi$ and cluster mass is difficult to characterize. Therefore, an unbiased SZE significance $\zeta$ is introduced, which is the signal-to-noise at the true, underlying cluster position and filter scale \citep{vanderlinde10}.
For $\zeta>2$, the relationship between $\xi$ and $\zeta$ is given by
\begin{equation} \label{eq:xi2zeta}
\zeta = \sqrt{\langle\xi\rangle^2-3}.
\end{equation}
The unbiased significance $\zeta$ is related to mass $M_{500,\textrm c}$ by
\begin{equation}
\zeta = A_{\textrm{SZ}}\left(\frac{M_{500,c}}{3\times 10^{14}M_{\odot}h^{-1}}\right)^{B_{\textrm{SZ}}}\left(\frac{E(z)}{E(0.6)}\right)^{C_{\textrm{SZ}}}
\label{eq:scalrelSZ}
\end{equation}
where $A_{\textrm{SZ}}$ is the normalization, $B_{\textrm{SZ}}$ the mass slope, $C_{\textrm{SZ}}$ the redshift evolution parameter and $E(z)\equiv H(z)/H_0$.  An additional parameter $D_{\textrm{SZ}}$ describes the intrinsic scatter in $\zeta$ which is assumed to be log-normal and constant as a function of mass and redshift. The scaling parameters and the priors we adopt are summarized in Table~\ref{tab:LCDM}, and further discussed in Section~\ref{sec:SZEpriors}.


\begin{turnpage}
\begin{deluxetable*}{lcccccccccccc}[htb]
  \tablecaption{\LCDM\ constraints from SZE cluster number counts \Nxiz\ with mass calibration from \Yx\ and $\sigma_v$, CMB and additional cosmological probes.}
    \tablehead {  \colhead{Param.} & \colhead{Prior} & \colhead{\Nxiz} & \multicolumn{3}{c}{\Nxiz+BBN+$H_0$+} & \colhead{WMAP9} & \multicolumn{2}{c}{\SPTcl+WMAP9} & \colhead{\planck+WP} & \multicolumn{2}{c}{\SPTcl+\planck+WP}\\ 
    &&&\colhead{\Yx} & \colhead{$\sigma_v$}& \colhead{\Yx+$\sigma_v$} &&& \colhead{+BAO+SNIa} &&& \colhead{+BAO+SNIa}}
    \startdata
    \Asz\ & $6.24\pm1.87$ & $6.49^{+2.08}_{-1.89}$ &	$5.59^{+1.19}_{-1.69}$ & $4.38^{+1.05}_{-1.45}$ & $4.70^{+0.82}_{-1.24}$&\nodata & $3.79^{+0.57}_{-0.63}$ & $3.47\pm0.48$ & \nodata & $3.27\pm0.35$ & $3.22\pm0.30$\\[6pt]
    \Bsz\ & $1.33\pm 0.266$ & $1.54\pm0.16$ &	$1.56\pm0.13$ &	$1.65\pm0.14$ &		$1.58\pm0.12$ &		\nodata	& $1.47\pm0.11$ & $1.48\pm0.11$ & \nodata & $1.49\pm0.11$ & $1.49\pm0.11$\\[6pt]
    \Csz\ & $0.83\pm 0.415$ & $0.75\pm0.39$ &	$0.82\pm0.35$ &	$0.92\pm0.37$ &		$0.91\pm0.35$ &		\nodata	& $0.40\pm0.23$ & $0.44\pm0.23$ & \nodata & $0.44\pm0.21$ & $0.49\pm0.22$\\[6pt]
    \Dsz\ & $0.24\pm 0.16$ & $0.32\pm0.16$ &	$0.28\pm0.11$ &	$0.24^{+0.11}_{-0.14}$ &	$0.26\pm0.10$ &		\nodata	& $0.25\pm0.10$ & $0.27\pm0.10$ & \nodata & $0.25\pm0.05$ & $0.26\pm0.05$\\[6pt]
    \tableline\\
    $A_\textrm X$ & $5.77 \pm 0.56$ &\nodata	& $5.40\pm0.56$ &\nodata& $5.76\pm0.50$ & 	\nodata & $5.79\pm0.43$ & $5.94\pm0.43$	& \nodata & $6.10\pm0.42$ & $6.13\pm0.40$\\[6pt]
    $B_\textrm X$ & $0.57 \pm 0.03$&\nodata	& $0.547\pm0.030$ &\nodata& $0.545\pm0.030$ & 	\nodata & $0.548\pm0.029$ & $0.549\pm0.029$	& \nodata & $0.546\pm0.029$ & $0.546\pm0.029$\\[6pt]
    $C_\textrm X$ & $-0.40 \pm 0.20$&\nodata	& $-0.37\pm0.18$ &\nodata& $-0.28\pm0.17$ & \nodata &  $-0.24\pm0.17$ & $-0.21\pm0.17$	& \nodata & $-0.17\pm0.16$ & $-0.16\pm0.16$\\[6pt]
    $D_\textrm X$ & $0.12 \pm 0.08$	&\nodata& $0.15\pm0.07$ &\nodata& $0.15\pm0.07$ & 	\nodata & $0.14\pm0.07$ & $0.14\pm0.07$	& \nodata & $0.14\pm0.07$ & $0.14\pm0.07$\\[6pt]
        \tableline\\
    \Adisp\ \tablenote{The units of \Adisp\ are km s$^{-1}$.} & $939 \pm 47$	&\nodata&\nodata& $971^{+47}_{-43}$ & $984\pm39$ &			\nodata & $973\pm35$ & $961\pm35$	& \nodata & $948\pm34$ & $946\pm33$\\[6pt]
    $B_{\sigma_v}$ & $2.91 \pm 0.15$&\nodata&\nodata& $2.91\pm0.16$ & $2.92\pm0.16$ & 	\nodata & $2.92\pm0.15$ & $2.92\pm0.16$	& \nodata & $2.92\pm0.16$ & $2.91\pm0.16$\\[6pt]
    $C_{\sigma_v}$ & $0.33 \pm 0.02$&\nodata&\nodata& $0.330\pm0.021$ & $0.331\pm0.021$ & 	\nodata & $0.329\pm0.021$ & $0.329\pm0.020$	& \nodata & $0.327\pm0.021$ & $0.328\pm0.020$\\[6pt]
    $D_{\sigma_v0}$ & $0.2\pm 0.04$	&\nodata&\nodata& $0.176\pm0.030$ & $0.176\pm0.030$ & \nodata & $0.176\pm0.028$ & $0.174\pm0.030$	& \nodata & $0.175\pm0.029$ & $0.175\pm0.029$\\[6pt]
    $D_{\sigma_v \textrm N}$ & $3\pm 0.6$	&\nodata&\nodata& $2.93\pm0.56$ & $2.92\pm0.56$ & \nodata & $2.92\pm 0.56$ & $2.93\pm0.54$	& \nodata & $2.93\pm0.54$ & $2.93\pm0.54$\\[6pt]
        \tableline\\
    $H_0$ \tablenote{The units of the Hubble constant $H_0$ are km s$^{-1}$ Mpc$^{-1}$.} & \nodata \tablenote{We apply a prior $H_0=73.8\pm2.4$~km~s$^{-1}$~Mpc$^{-1}$ when no CMB data are included in the fit.} & $73.5\pm2.4$ & $73.2\pm2.5$ & $73.4\pm2.4$ & $73.2\pm2.6$ & $70.0\pm2.4$ & $70.1\pm1.7$ & $68.6\pm1.0$ & $67.6\pm1.2$ & $68.6\pm1.1$ & $68.3\pm0.8$\\[6pt]
    \Om\ &\nodata & $0.39^{+0.07}_{-0.13}$ & $0.41^{+0.07}_{-0.14}$ & $0.45^{+0.09}_{-0.16}$ & $0.44^{+0.07}_{-0.15}$ & $0.281\pm0.028$ & $0.276\pm0.018$ & $0.292\pm0.011$ & $0.310\pm0.017$ & $0.297\pm0.014$ & $0.299\pm0.009$\\[6pt]
    $\sigma_8$ &\nodata& $0.67\pm 0.07$ & $0.69\pm0.06$ & $0.72\pm0.07$ & $0.71\pm0.06$ & $0.825\pm0.027$ & $0.812\pm0.017$ & $0.816\pm0.016$ & $0.841\pm0.013$ & $0.828\pm0.011$ & $0.829\pm0.011$\\[6pt]
  \multicolumn{2}{l}{$\sigma_8\left( \frac{\Omega_\textrm m}{0.27}\right)^{0.3}$ \tablenote{The uncertainty on \widthparam\ reflects the width of the likelihood contour in the direction orthogonal to the cluster degeneracy in the \Om-$\sigma_8$ plane.} \nodata} & $0.741\pm0.064$ & $0.774\pm0.040$ & $0.831\pm0.052$ & $0.809\pm0.036$ & $0.835\pm0.051$ & $0.817\pm0.027$ & $0.835\pm0.022$ & $0.877\pm0.024$ & $0.852\pm0.020$ & $0.855\pm0.016$
   \enddata
\tablecomments{\Nxiz\ denotes the cluster sample without additional mass calibration information; \SPTcl\ contains the clusters with the mass calibration data from X-ray \Yx\ and velocity dispersion $\sigma_v$. The priors are Gaussian as discussed in Section~\ref{sec:priors}. The scalar spectral index $n_\textrm s$, the reionization optical depth $\tau$, the baryon density $\Omega_\textrm b$, and the \planck\ nuisance parameters are not shown in this table but are included in the analysis and marginalized out. We fix $\tau = 0.089$ when no CMB data are included in the fit.}
  \label{tab:LCDM}
\end{deluxetable*}
\end{turnpage}


We use SPT-selected clusters for the cosmological cluster number count and mass calibration analysis, described in Section~\ref{sec:method}.
For the number counts, we use a cluster sample identical to the one used in \cite{reichardt13}. This sample uses data from the first 720~deg$^2$ of the SPT-SZ survey and is restricted to $\xi>5$ and redshift $z>0.3$; it contains 100 cluster candidates.
No optical counterparts were found for six of these SZE detections; we discuss their treatment in the analysis in Section~\ref{sec:unconfirmed}.
The SPT-SZ 720~deg$^2$ survey comprises 5~fields with different depths which are accounted for by rescaling the SPT $\zeta$-mass relation normalization \Asz\ for each field \citep{reichardt13}.
Our mass calibration data consists of a sub-sample of 64 SPT clusters with additional X-ray and/or spectroscopic follow-up data, as described in Section~\ref{sec:optspec} and \ref{sec:Xrayobservations}.
Twenty-two clusters with velocity dispersion \sigmav\ measurements lie outside the SPT-SZ 720~deg$^2$ survey. The depths of these fields and the corresponding scaling factors for \Asz\ will be presented elsewhere together with the analysis of the full 2500~deg$^2$ survey catalog (de Haan et al. in preparation). These scaling factors are all between $1.08-1.27$ with a median value of 1.17.

\subsection{Optical and Near-Infrared Imaging}
The galaxy clusters analyzed here have been followed up in optical and near infrared in the context of the SPT follow-up program, as described in \citet{song12b}, to which we refer the reader for details of the strategy and data reduction.
Briefly, the SPT strategy is to target all galaxy clusters detected at SZE significance $\xi >4.5$ for multiband imaging in order to identify counterparts to the SZE signal and obtain photometric redshifts.
We also obtain \spitzer/IRAC near-infrared imaging for every cluster with SZE significance $\xi>4.8$, and we target those systems at lower $\xi$ which are not optically confirmed or have a redshift above 0.9 with ground based near infrared imaging using the NEWFIRM imager on the CTIO Blanco 4~m telescope.

\subsection{Optical Spectroscopy} \label{sec:optspec}

We use follow-up optical spectroscopy to measure the velocity dispersion \sigmav\ of 63 clusters. Of these, 53 were observed by the SPT team \citep{ruel13} and 10 have data taken from the literature \citep{barrena02,buckleygeer11,sifon13}. In \cite{ruel13}, four~additional clusters with spectroscopic data are listed, but we choose not to include them in our analysis as they are all at relatively low redshifts below $z<0.1$ where the SZE mass-observable scaling relation we adopt is likely not valid. The lowest redshift cluster entering our mass calibration analysis is SPT-CL~J2300-5331 at $z=0.2623$.

Our own data come from a total observation time of $\sim70$~h on the largest optical telescopes (Gemini South, Magellan, and VLT) in the southern hemisphere; we specifically designed these observations to deliver the data needed for this velocity dispersion mass calibration study.  We obtained low-resolution ($R\simeq 300$) spectra using several different instruments: GMOS\footnote{http://www.gemini.edu/node/10625} on Gemini South, FORS2 \citep{appenzeller98} on VLT Antu, LDSS3 on Magellan Clay and IMACS/Gladders Image-Slicing Multislit Option (GISMO\footnote{http://www.lco.cl/telescopes-information/magellan/\\instruments/imacs/gismo/gismoquickmanual.pdf}) on Magellan Baade.

Apart from early longslit spectroscopy using the Magellan LDSS3 spectrograph on a few SPT clusters, the general strategy is to design two masks per cluster for multi-object spectroscopy to get a final average number of 25 member galaxy redshifts per cluster.
We typically obtained deep ($m^{\star}+1$) pre-imaging in $i'$-band for spectroscopic observation to (1) accurately localize galaxies to build masks for multi-object spectroscopy, and (2) identify possible giant arcs around cluster cores. This deep pre-imaging is used together with existing shallower optical imaging and near infrared photometry, where available, to select galaxy cluster members along the red sequence. We refer the reader to \citet{ruel13} for a detailed description of the cluster member selection and the data reduction.

\subsection{X-ray Observations and \Yx\ Scaling Relation Parametrization}
\label{sec:Xrayobservations}

Sixteen clusters of our sample have been observed in X-ray using either \chandra\ or \xmm.  The derived properties of 15 of these clusters are published in \citet{andersson11}. This sub-sample corresponds to the highest SZE significance clusters in the first 178~deg$^2$ of the SPT-SZ survey that lie at $z\gtrsim0.3$. We obtained \chandra\ observations of SPT-CL~J2106-5844 in a separate program whose results are published elsewhere \citep{foley11}.  All of these observations have $>1500$ source photons within $0.5\times r_{500}$ and in the 0.5-7.0~keV energy band.  X-ray observations are used to derive the intracluster medium temperature $T_\textrm X$ and the gas mass $M_\textrm g$. For a detailed description of the data reduction method, we refer the reader to \citet{andersson11}. Note that there is a calibration offset between temperature measurements from the two satellites \citep{schellenberger14}. For our analysis, we adopt priors on the \Yx-mass relation that come from an analysis of \chandra\ data. Given that only 2/16 systems in this study rely on \xmm\ data, and the amplitude of the calibration offset is $\sim$30\% in temperature for these massive clusters, we expect an overall temperature bias of $\sim$4\%, corresponding to a $\sim$2\% bias in our mass scale, assuming that the \chandra-derived temperatures are unbiased. Given that this is much smaller than the systematic uncertainty in our \Yx-mass calibration, we neglect any cross-calibration.

Following \cite{benson13} we rely on the X-ray observable $Y_\textrm X\equiv M_\textrm g T_\textrm X$. For the cosmological analysis performed in this work we need to evaluate \Yx\ as a function of cosmology and scaling relation parameters. In practice, for a given set of cosmological and scaling relation parameters, we iteratively fit for $r_{500}$ and $Y_\textrm X(r)$ which is then used to estimate the cluster mass.

We adopt a calibrated scaling relation derived from hydrostatic masses at low redshifts \citep{vikhlinin09a}:
\begin{equation} \label{eq:scalrelx}
  \frac{M_{500,c}}{10^{14}M_{\odot}}  = A_\textrm X h^{1/2} \left( \frac{Y_\textrm X}{3\times
      10^{14}M_{\odot} \mathrm{keV}}\right)^{B_\textrm X} E(z)^{C_\textrm X},
\end{equation}
where $A_\textrm X$ is the normalization, $B_\textrm X$ the slope and $C_\textrm X$ the redshift evolution parameter.  We assume an intrinsic log-normal scatter in \Yx\ denoted $D_\textrm X$ and an observational log-normal uncertainty for each cluster. The fiducial values and priors we adopt for the \Yx\ parameters are discussed in Section~\ref{sec:Yxpriors} and shown in Table~\ref{tab:LCDM}.

\subsection{External Cosmological Datasets}
In addition to our cluster sample, we include external cosmological datasets such as measurements of the CMB anisotropy power spectrum, the baryon acoustic oscillations (BAO), Type~Ia Supernovae (SNIa), the Hubble constant ($H_0$), and Big Bang nucleosynthesis (BBN). We use these abbreviations when including the datasets in the analysis. We refer to the SPT SZE cluster sample without the follow-up mass information as \Nxiz (which stands for the distribution of the clusters in $\xi$-$z$ space), and we refer to the full cluster sample with mass measurements from \sigmav\ and \Yx\ as \SPTcl.

We include measurements of the CMB anisotropy power spectrum from two all-sky surveys. We use data from the \textit{Wilkinson Microwave Anisotropy Probe} \citep[WMAP, 9-year release;][]{hinshaw13} and data from the \planck\ satellite \cite[1-year release, including WMAP polarization data (WP);][]{planck13-25,planck13-26}.
The BAO constraints are applied as three measurements:  $D_\textrm V(z=0.106)=457\pm27 \,\textrm{Mpc}$ \citep{beutler11}, $D_\textrm V(z=0.35)/r_\textrm s=8.88\pm0.17$ \citep{padmanabhan12}, and $D_\textrm V(z=0.57)/r_\textrm s=13.67\pm0.22$ \citep{anderson12}; $r_\textrm s$ is the comoving sound horizon at the baryon drag epoch, $D_\textrm V(z)\equiv [(1+z)^2D_\textrm A^2(z)cz/H(z)]^{1/3}$, and $D_\textrm A$ is the angular diameter distance.
We include distance measurements coming from Type Ia supernovae using the Union2.1 compilation of 580~SNe \citep{suzuki12b}.  We adopt a Gaussian prior on the Hubble constant $H_0 = 73.8\pm2.4 \textrm{~km~s$^{-1}$~Mpc$^{-1}$}$ from the low-redshift measurements from the \textit{Hubble Space Telescope} \citep{riess11}. Finally we use a BBN prior from measurements of the abundance of $^4$He and deuterium which we include as a Gaussian prior $\Omega_\textrm bh^2 = 0.022\pm0.002$ \citep{kirkman03}.
Note that both the BBN and $H_0$ priors are only applied when analyzing the cluster samples without CMB data.


\section{Velocity Dispersions \sigmav\ as \\ Mass Calibrators}
\label{sec:velocitydispersion}

Multiple studies highlight the fact that the line-of-sight velocity dispersion of galaxies within clusters may be used to measure galaxy cluster masses \citep[e.g.,][]{biviano06,evrard08,white10,munari13,saro13}. The motivation to use velocity dispersions as a mass probe for galaxy clusters stems from the fact that the galaxy dynamics are unaffected by the complex physics of the intracluster medium. Therefore, the dominant source of scatter and bias in the $\sigma_v$-mass scaling relation is related to gravitational dynamics of subhalos, an effect that can be studied using high-resolution $N$-body simulations. As we will discuss in Section~\ref{sec:disppriors}, the systematic floor on dynamical mass, which is due to uncertainties in modeling the velocity bias, is currently of the order of 15\% in mass (equivalent to 5\% in \sigmav).

\citet{saro13} used the publicly available galaxy catalogs produced with the semi-analytic model \citep{delucia07a} from the Millennium simulation \citep{springel05} to precisely characterize the $\sigma_v$-mass scaling relation as a function of parameters such as redshift, number of selected red-sequence galaxy cluster members and aperture size centered on the cluster. Their approach provides a mapping between $\sigma_v$ and cluster mass that includes the effects of galaxy selection, departures from equilibrium and sample size, all of which can be used to interpret the velocity dispersions available for our SPT clusters. There are two important, but opposing effects that may lead to a potential bias: (1) dynamical friction, which biases the velocity dispersion low, and (2) interlopers, which for our selection tend to bias dispersions high. For our selection approach, these contributions effectively cancel, producing no net bias. The intrinsic scatter on an individual dynamical mass is typically 80\% due to the random projection of the velocity ellipsoid along the line of sight and interlopers in the calculation of velocity dispersion.

Given the large mass uncertainty associated with the dispersion from an individual cluster, we use a large ensemble of dispersion measurements for our mass calibration analysis. Within this context, we should be able to constrain the normalization \Asz\ of the SZE $\xi$-mass relation to a level where it is dominated by the 15\% systematic uncertainty in the dispersion mass estimates. However, because the intrinsic scatter in the velocity dispersion scaling relation is much larger than the scatter in the SZE $\xi$-mass scaling relation, we do not expect to improve our constraints on the scatter of the SZE $\xi$-mass scaling relation using velocity dispersions.

We assume the scatter in \sigmav\ to be uncorrelated with the scatter in SZE.
In principle, cluster triaxiality might induce such a correlation; however, for our sample, the intrinsic scatter in \sigmav\ is dominated by the effect of interlopers, which do not affect the SZE signal.

We adopt the mass-observable scaling relation for velocity dispersions \sigmav\ presented in \cite{saro13}:
\begin{equation}
M_{200,c} = \left( \frac{\sigma_v}{A_{\sigma_v}h_{70}(z)^{C_{\sigma_v}}} \right)^{B_{\sigma_v}}  10^{15} M_{\odot}
\end{equation}
where $M_{200,c}$ is the mass expressed relative to the critical density, \Adisp\ is the normalization, $B_{\sigma_v}$ the slope, and $C_{\sigma_v}$ the redshift evolution parameter.
We express the scatter in \sigmav\ as a function of $N_\mathrm{gal}$, the number of spectroscopically observed cluster galaxies. The scatter is described by a log-normal distribution of width
\begin{equation}
D_{\sigma_v} = D_{\sigma_v0} + D_{\sigma_v \textrm N}/N_\mathrm{gal}
\end{equation}
where $D_{\sigma_v0}$ and $D_{\sigma_v \textrm N}$ are two parameters extracted from the simulations. Given that the typical number of spectroscopically observed galaxies is small for our sample, this dependency of the scatter on $N_\mathrm{gal}$ is important for our analysis.  The fiducial values and priors adopted for the parameters are discussed in Section~\ref{sec:disppriors} and shown in Table~\ref{tab:LCDM}.

Note that the SZE and X-ray mass scaling relations are defined in terms of $M_{500,c}$ whereas the dynamical mass is defined as $M_{200,c}$.  The mass conversion is performed using the NFW profile \citep{navarro97} and the \citet{duffy08} mass-concentration relation.


\section{Analysis Method}
\label{sec:method}

In this Section we introduce the likelihood model adopted for analyzing the data. When combining the cluster experiment with other cosmological probes, we multiply the individual likelihoods.
The multi-dimensional parameter fit varying all relevant cosmological and scaling relation parameters is performed using a Population Monte Carlo (PMC) algorithm as implemented in the CosmoPMC code \citep{kilbinger11}.  In contrast to the widely used Markov Chain Monte Carlo (MCMC) method, which explores the parameter space based on an acceptance-rejection algorithm, the PMC algorithm iteratively fits for the posterior distribution using samples of points (populations) in parameter space. This leads to a significant reduction of computational time as (1) the calculations of the likelihood at individual points in parameter space are independent and therefore can be computed in parallel and (2) the overall efficiency is higher than when using MCMC as there are no rejected points. For a detailed description of the PMC algorithm and its comparison with MCMC see e.g., \citet{Wraith:2009if}.

When analyzing the \SPTcl\ sample without CMB data we fit for up to 18 parameters: 4 SZE, 4 \Yx, 5 $\sigma_v$ scaling relation parameters, and 5 cosmological parameters ($\sigma_8$, \Om, $\Omega_\textrm b$, $H_0$, $n_\textrm s$); we fix the optical depth because it is not constrained by the data. When combining with the CMB dataset from WMAP we also include the optical depth $\tau$ as a free parameter in the fit; when analyzing \planck\ data we include further nuisance parameters.

We finally describe the priors that we adopt for each of the mass-observable scaling relations and explain how we tested our code using mock data.

\subsection{Likelihood Model}
\label{sec:likelihoodmodel}
The cluster number count analysis in the SZE observable $\xi$ can be separated from the additional mass calibration in an unbiased way. This approach allows for an easy comparison and combination of the different mass calibrators as we will discuss in Section~\ref{sec:methodcompare}. For a detailed derivation of our likelihood function, see Appendix.

\subsubsection{Cluster Mass Function} \label{sec:clustermassfunction}

At each point in the space of cosmological and scaling-relation parameters we use the Code for Anisotropies in the Microwave Background \citep[CAMB,][]{lewis00} to compute the matter power spectrum at 180 evenly spaced redshift bins between $0.2<z<2$. We then use the fitting function presented in \cite{tinker08} to calculate the cluster mass function $dN/dM$ for 500 mass bins evenly distributed in log-space between $10^{13.5}h^{-1}M_\odot\le M\le 10^{16}h^{-1} M_\odot$. This fitting function is accurate at the 5\% level across a mass range $10^{11}h^{-1}M_\odot\le M\le 10^{15}h^{-1} M_\odot$ and for redshifts $z \le 2.5$.

We move the mass function from its native mass and redshift space to the observable space in $\xi$-$z$:
\begin{eqnarray}
\label{eq:obsmassfunc}
\frac{dN(\xi,z | \vect{p})}{d\xi dz} &= \int dMdz\, \Theta(\xi - 5, z-0.3) \times \nonumber \\
& P(\xi | M, z, \vect{p}) \otimes \frac{dn(M,z|\vect{p})}{dM} \frac{dV(z)}{dz} 
\end{eqnarray}
where $dV/dz$ is the comoving volume within each redshift bin, $\vect{p}$ is a vector containing all scaling relation and cosmological parameters, and $\Theta$ is the Heaviside step function describing cluster selection in the SZE observable $\xi>5$, and observed redshift $z>0.3$.  The term $P(\xi| M, z, \vect{p})$ describes the relationship between mass and the SZE observable from the scaling relation (Equations \ref{eq:xi2zeta} and \ref{eq:scalrelSZ}), and contains both intrinsic and observational uncertainties. In practice, we convolve the mass function with this probability distribution.

Finally, the logarithm of the likelihood $\mathcal L$ for the observed cluster counts is computed following \cite{cash79}. After dividing up the observable space in small bins, the number of expected clusters in each bin is assumed to follow a Poisson distribution. With this the likelihood function is
\begin{equation}
\label{eq:numbercountlikelihood}
\ln \mathcal{L}(\vect{p}) = \sum_i  \ln \frac{dN(\xi_i,z_i | \vect{p})}{d\xi  dz} - \int \frac{dN(\xi,z | \vect{p})}{d\xi dz} d\xi  dz,
\end{equation}
up to a constant offset, and where $i$ runs over all clusters in the catalog. For clusters without spectroscopic data, we integrate the model over redshift weighting with a Gaussian whose central value and width correspond to the cluster's photometric redshift measurement.

The 720~deg$^2$ survey area contains five fields of different depths, see Section~\ref{sec:SZdata}. In practice, we perform the above calculation for each field rescaling \Asz\ with the corresponding factor, and sum the resulting log likelihoods.

\subsubsection{Mass Calibration}
\label{sec:masscalibration}

For each cluster in our sample containing additional mass calibration information from X-ray and/or velocity dispersions, we include the \Yx\ or \sigmav\ measurement as follows: At every point in cosmological and scaling relation parameter space $\vect{p}$, we calculate the probability distribution $P(M | \xi,z,\vect{p})$ for each cluster mass, given that the cluster has a measured significance $\xi$ and redshift $z$:
\begin{equation}
P(M | \xi,z,\vect{p}) \propto P(\xi | M,z,\vect{p}) P(M|z,\vect{p}).
\end{equation}
In practice, we calculate the probability distribution $P(\xi | M,z,\vect{p})$ from the SZE scaling relation (Equations~\ref{eq:xi2zeta} and \ref{eq:scalrelSZ}) taking both intrinsic and observational scatter into account, and weight by the mass function $P(M|z,\vect{p})$, thereby correcting for Eddington bias. We then calculate the expected probability distribution in the follow-up observable(s) which we here call $\mathcal O$ for simplicity:
\begin{equation} \label{eq:P_O_xi}
P(\mathcal O|\xi,z,\vect{p}) = \int dM \,P(\mathcal O|M,z,\vect{p})P(M|\xi,z,\vect{p}).
\end{equation}
The term $P(\mathcal O|M,z,\vect{p})$ contains the intrinsic scatter and observational uncertainties in the follow-up observable.  We assume the intrinsic scatter in the SZE scaling relation and the follow-up measurements to be uncorrelated.  For each cluster in the mass calibration sample, we compare the predicted $P(\mathcal O|\xi,z,\vect{p})$ with the actual measurement and extract the probability of consistency. Finally, we sum the log-likelihoods for all these clusters and add the result to the number count likelihood (Equation \ref{eq:numbercountlikelihood}).

It is important that any cosmological dependence of the mass calibration observations be accounted for.  In the case of a single velocity dispersion $\sigma_v$, the measurement comes from the combination of redshift measurements from a sample of cluster galaxies; the cosmological sensitivity, if any, is subtle.  On the other hand, the X-ray observable \Yx\ is calculated from the measured temperature and gas mass within  $r_{500}$, and the limiting radius and the gas mass are both cosmology dependent.  Therefore, \Yx\ has to be extracted from the observations for each set of cosmological and scaling relation parameters as described in Section~\ref{sec:Xrayobservations}.

\subsubsection{Unconfirmed Cluster Candidates}
\label{sec:unconfirmed}
Out of the 100 cluster candidates in the survey, 6 detections could not be confirmed by the optical follow-up and were assigned lower redshift limits based on the depth of the imaging data \citep{song12b}. In addition, each of these unconfirmed candidates has some probability of being a noise fluctuation.

Our treatment of these candidates takes into account the false detection rate at the detection signal-to-noise as well as the expected number of clusters exceeding the lower redshift bound of the candidate as predicted by the cluster mass function. We calculate the probability of a candidate $i$ to be a true cluster according to
\begin{equation}
P_\textrm{true}^i = \frac{N_\textrm{expected}(\xi^i, z_\textrm{low}^i| \vect{p})}{N_\textrm{expected}(\xi^i, z_\textrm{low}^i| \vect{p})+N_\textrm{false detect}(\xi^i)}
\end{equation}
where the number of clusters $N_\textrm{expected}$ above some lower redshift limit is given by $\int_{z_\textrm{low}^i}^\infty N(\xi^i, z|\vect{p})dz$. The expected number of false detections as a function of $\xi$ has been estimated from simulations and cross-checked against direct follow-up and is assumed to be redshift independent \citep{song12b,reichardt13}.

In the cosmological analysis, each of the unconfirmed candidates is treated like an actual cluster but weighted with its $P_\textrm{true}^i$. However, the specific treatment of the unconfirmed candidates has little effect on the cosmological and scaling relation parameters; for example, simply removing these candidates from the catalog leads to negligible changes in the results.

\subsection{Discussion of the Analysis Method}
\label{sec:methodcompare}

In previous SPT cluster cosmology studies, we have used a somewhat different method.  In that method the expected number density of clusters as a function of $\xi$, \Yx, and $z$ is calculated on a three-dimensional grid. The likelihood is evaluated by comparing this prediction to the cluster sample in a way analogous to Equation~\ref{eq:numbercountlikelihood}. For clusters without \Yx\ data the likelihood is integrated over the full range of \Yx\ \citep{benson13}.

As we show in the Appendix, the method we employ in the current analysis is mathematically equivalent to this other method; here we assume uncorrelated scatter.
For the current application, where we have \sigmav\ and \Yx\ follow-up measurements, we do not work in the four-dimensional $\xi$-\Yx-\sigmav-$z$-space, but rather we treat the number count part of the likelihood in its $\xi$-$z$-space, and the mass calibration part of the likelihood $P(\mathcal O|\xi,z,\vect{p})$ separately. The results obtained with this analysis method do not show any sign of biases when tested against different sets of mock data (see Section~\ref{sec:testmock}).  This method is convenient when analyzing a cluster sample with multiple different mass observables where only a fraction of the clusters have those observables.  In the limit where every cluster in the survey has the same follow-up mass measurements, the likelihood presented and used in our previous analyses \citep{benson13,reichardt13} would be more computationally efficient.

\subsection{Priors Used in the Analysis}
\label{sec:priors}

We present the priors used in our analysis and discuss their motivation. All priors are also listed in the first column of Table \ref{tab:LCDM}.

\subsubsection{Priors on SZE $\xi$-mass Scaling Relation Parameters}
\label{sec:SZEpriors}

The SZE scaling relation parameters were estimated from simulations of the SZE sky of about 4000~deg$^2$ in size \citep{reichardt13}.
We adopt 30\%, 20\%, 50\% Gaussian uncertainties on \Asz, \Bsz, and \Csz, respectively \citep[e.g.,][]{vanderlinde10}. For the scatter \Dsz, we adopt a conservative 67\% uncertainty \citep{benson13, reichardt13}.

\subsubsection{Priors on \Yx-mass Scaling Relation Parameters}
\label{sec:Yxpriors}

The priors used in the X-ray scaling relation parameters are motivated by published constraints from X-ray measurements and simulations.  The absolute mass scale of the \Yx-mass scaling relation has been calibrated using hydrostatic mass estimates of a sample of 17 low-redshift ($z<0.3$) relaxed clusters \citep{vikhlinin09a}. Simulations were used to estimate an upper limit of 4\% on the systematic offset in the \Yx-mass relation between relaxed and unrelaxed clusters \citep{kravtsov06a} .  Also, simulations predict that biases in hydrostatic mass estimates are less for relaxed clusters and are of the order of 15\% \citep{nagai07}.  Therefore, the \Yx-mass relationship calibrated from hydrostatic mass of a sample of relaxed clusters should be in principle applicable to less relaxed systems.

We adopt the best-fit value of $A_\textrm X = 5.77\pm 0.20$ for the normalization and $B_\textrm X = 0.57 \pm 0.03$ for the slope where uncertainties are statistical only \citep{vikhlinin09a}.  The systematic uncertainty on $A_\textrm X$ was determined by comparing to weak-lensing mass estimates for a sample of 10 low-redshift clusters \citep{hoekstra07} . The derived $1\sigma$ systematic uncertainty is 9\% on the \chandra\ mass calibration. Adding this in quadrature to the statistical uncertainty yields the Gaussian prior $A_\textrm X =5.77\pm 0.56$ we use in this study.

For the redshift evolution parameter, we assume a Gaussian prior $C_\textrm X = -0.4\pm 0.2$. The 50\% uncertainty is motivated by simulations \citep{kravtsov06a} and matches the prior used in the hydrostatic calibration analysis \citep{vikhlinin09a}.

We apply a Gaussian prior $D_\textrm X = 0.12\pm 0.08$ on the log-normal intrinsic scatter. The central value of the prior is chosen to be consistent with simulations \citep[e.g.,][]{kravtsov06a}, while the uncertainty is chosen to encompass the range found in simulations and in measured values in the literature \citep{vikhlinin09a, mantz10b}.

\subsubsection{Priors on $\sigma_v$-mass Scaling Relation Parameters}
\label{sec:disppriors}

The statistical uncertainty on the normalization \Adisp\ of the relation is of the order of $0.06\%$ \citep{saro13}.  However, there is a systematic uncertainty associated with the poorly determined galaxy velocity bias $b$, and this has been the focus of multiple investigations. Remember that $b=1$ means no bias. For example, from the analysis of the Millennium simulation \citep{springel05}, a weak velocity bias of $1.02$ is claimed \citep{faltenbacher06}, while \citet{biviano06} derive a bias of $0.95$ using gas dynamic simulations \citep{borgani04}.  Based on the comparison of different simulations, \citet{evrard08} estimates a bias of $1.00\pm 0.05$, and \citet{white10} derives a value $\sim1.06$ from their own $N$-body simulation.  In more recent studies comparing different simulations, \cite{wu13} and \cite{gifford13} find a spread in velocity bias of the order of 10\%.  Taking into account these different results, we adopt a Gaussian 5\% prior on the normalization of the scaling relation centered at the value given by \citet{saro13}: $A_{\sigma_v} = 939\pm 47$~km~s$^{-1}$.  This corresponds to a 15\% systematic uncertainty floor in the velocity dispersion mass estimates used in our analysis.  We expect future studies to help in providing more accurate estimations of the velocity bias.

In our recent presentation of the velocity dispersion data on the SPT cluster sample \citep{ruel13} we note a 10\% offset in the dispersion normalization of the dataset as compared to the predicted dispersions \citep{saro13} when using the previously published SPT cluster masses \citep{reichardt13}.  Stated in another way, this offset is an indication that if the dispersions were used for mass calibration, then they would lead to a change in the mass scale of the SPT cluster sample. This expectation is confirmed in the results presented below (see Section~\ref{sec:clustersonly}).

\cite{saro13} find the statistical uncertainties for the slope $B_{\sigma_v}$ and the evolution term $C_{\sigma_v}$ to be $\mathcal O(10^{-4})$ and $\mathcal O(10^{-3})$, respectively, and hence completely negligible. However, these results do not include potential systematic uncertainties. We adopt conservative 5\% Gaussian uncertainties on both parameters and apply $B_{\sigma_v} = 2.91\pm0.15$ and $C_{\sigma_v} =0.33\pm0.02$. We confirm that the width of those priors plays a negligible role in our analysis by tightening both priors to the levels of the statistical uncertainties quoted above; the results on all other parameters remain essentially unchanged.

The effect of interlopers is the dominant contribution to the intrinsic scatter \citep{saro13} and we assume a 20\% uncertainty on the scatter normalization $D_{\sigma0} = 0.2\pm0.04$ as well as a 20\% uncertainty on its dependence on the number of observed galaxies $D_{\sigma N}=3\pm0.6$.
The results from our observed velocity dispersion sample support this approach; we measure the scatter in the observed sample to be $D_{\sigma_v}=0.31\pm0.03$ \citep{ruel13}.  In the present analysis we use a parametrization of the scatter that includes the number of spectroscopically observed galaxies (see Section~\ref{sec:velocitydispersion}). For the typical number of observed galaxies in our sample $\langle N_\textrm{gal}\rangle=25$, we model the scatter to be $D_{\sigma_v} ({N_\textrm{gal}=25})=0.32$, which is in very good agreement with the direct measurement.

\subsubsection{Additional Priors on Cosmological Parameters}
\label{sec:additionalpriors}
Galaxy clusters are not sensitive to all cosmological parameters.
Therefore, when not including the CMB dataset in a cosmological analysis, we fix the optical depth at reionization to the WMAP9 best-fit value $\tau = 0.089$ and we adopt a Gaussian prior on the spectral index $n_\textrm s=0.972\pm0.013$ representing the WMAP9 result.

\subsection{Validation of the Analysis Tool using Mock Data} \label{sec:testmock}

We validate the analysis method using simulated data. In a first step we test the number count part in SZE significance and redshift space using simulated cluster catalogs that match the SPT data but contain orders of magnitude more clusters; our goal here is to minimize statistical noise so as to resolve possible systematics in the analysis at a level far below the statistical noise in our real sample. Our mock generator produces clusters in mass-redshift space, converts the cluster masses to the SZE observable $\xi$ using Equations \ref{eq:xi2zeta} and \ref{eq:scalrelSZ} with log-normal and normal scatter, respectively, and then applies the survey selection. The crucial part of the analysis - that is the conversion from mass to observable - is thereby computed differently than in the likelihood code we use to explore cosmological parameter space.

We generate large catalogs using different sets of input values and obtain samples containing on the order of $10^4$ clusters. We then run our analysis pipeline on the mock data using priors equivalent to the ones listed in Table \ref{tab:LCDM}; our tests show that we are able to recover the input values to within $1\sigma$ statistical uncertainties, verifying that there are no biases in our codes at a level well below the statistical noise in our real cluster ensemble.

We further analyzed mock catalogs produced using the analysis pipeline used in our previous analyses \citep{benson13,reichardt13}, recovering the input parameters at the $1\sigma$ statistical level.  To test the mass calibration module, we use a subset of 500 clusters drawn from the SZE mock catalog described above and additionally convert the cluster masses to X-ray \Yx\ and velocity dispersion $\sigma_v$ measurements. We then run our analysis code on the mass calibration part alone, that is without using the number count information and use \Yx\ and/or $\sigma_v$, showing that we are able to recover the input values.  Finally we confirm that the combination of number counts and mass calibration produces unbiased results by combining the SZE mock catalog with the X-ray and spectroscopic cluster mass observables.  These tests give us confidence that our code is producing unbiased constraints.


\section{Results}
\label{sec:results}

In this section, we present the results of our mass calibration and cosmological analysis. As we discuss in detail, the constraints obtained using \sigmav\ mass calibration are statistically consistent with those we obtain using \Yx, but the dispersions prefer higher cluster masses. 
Assuming a flat \LCDM\ cosmology, we compare the constraints obtained from the SPT galaxy clusters and mass calibration with independent cosmological constraints from CMB anisotropies, and finally combine the datasets in order to obtain tighter cosmological constraints.
We then use the combined datasets to constrain extensions of the standard cosmological model in which the Dark Energy equation of state or the sum of neutrino masses are allowed to vary.
Finally, we present the first SPT result on the cosmological growth of structure.


\subsection{Using $\sigma_v$ and  \Yx\ as Mass Calibrators}
\label{sec:clustersonly}

In Table~\ref{tab:LCDM}, we present the results of the analysis of the SPT-SZ survey cluster sample and its mass calibration assuming a flat \LCDM\ model. For now we do not include CMB, BAO, or SNIa data, because we first wish to isolate the galaxy cluster constraints and the impact of the mass calibration data. However, we include the BBN and $H_0$ priors, because not all parameters are well constrained by the cluster data.  

We present results using the SPT cluster sample \Nxiz\ only, \Nxiz\ with \Yx\ data, \Nxiz\ with \sigmav\ data, and \Nxiz\ with both \Yx\ and \sigmav.  It is clear that the additional mass information from \sigmav\ or \Yx\ help in improving the results obtained from \Nxiz\ only. The constraints on the SZE scaling relation normalization \Asz, the scatter in that relation \Dsz, and the cosmological parameter combination \widthparam\ tighten. The uncertainty on this parameter reflects the width of the likelihood distribution in \Om-$\sigma_8$ space in the direction orthogonal to the cluster degeneracy (see Figure~\ref{fig:Omegam-sigma8}).

There is agreement between the results obtained using the mass calibrators \sigmav\ or \Yx, which provides an indication that both methods are reliable and that systematics are under control.  The normalization \Asz\ decreases by 22\% when replacing the \Yx\ calibration dataset with the $\sigma_v$ dataset. Due to the skewness of the probability distributions with tails towards larger values, the constraints on \Asz\ from \sigmav\ and \Yx\ measurements have significant overlap, with the \Yx-favored value displaced $1.15\sigma$ from the result obtained from \sigmav\ (see also Figure~\ref{fig:Asz}).  The constraints on the slope \Bsz, the redshift evolution parameter \Csz, as well as the scatter \Dsz\ are not much affected by the choice of the mass calibrator.
We note that the \Yx\ scaling relation is calibrated by observations at $z\sim0.3$ which is extrapolated to higher redshifts using priors motivated by simulations, whereas the $\sigma_v$ scaling relation is calibrated to simulations over the full redshift range.  In terms of the cosmological results, both follow-up methods perform similarly in constraining the fully marginalized values for \Om\ and $\sigma_8$. However, the \Yx\ calibration does better in constraining \widthparam.

Our constraints using SPT clusters with mass calibration from X-ray \Yx\ only are comparable with previously published results from nearly the same cluster sample \citep{reichardt13}. Note that the X-ray sample used here contains measurements of \Yx\ for two additional clusters (see Section~\ref{sec:Xrayobservations}). We recover almost identical constraints on the SZE and X-ray scaling relation parameters. However, in the \Om-$\sigma_8$ plane, the constraints presented here extend further along the degeneracy direction towards higher values of \Om. This difference is due to a prior on the power spectrum normalization $\ln(10^{-10} A_s)=[2.3,4]$ that was narrow enough to affect the cosmological constraints in \citet{reichardt13}; we fit for $\sigma_8$ in the range $[0.4, 1.2]$ which is much broader than the recovered probability distribution and hence our choice of prior does not affect our results.

We estimate the effect of potentially larger galaxy velocity bias (see discussion in Section~\ref{sec:velocitydispersion} and~\ref{sec:disppriors}) by loosening our prior on \Adisp\ from the 5\% recommended by \cite{saro13} to 10\% when analyzing the \Nxiz+$\sigma_v$+BBN+$H_0$ data. There is a broadening of the uncertainty on \Asz\ by 25\%, and a $\sim0.3\sigma$ shift to a higher value. The constraint on \widthparam degrades by 14\% and shifts only by a negligible amount. In addition, we examine the impact of tightening the prior on \Adisp\ to 1\%. In this case, we observe improvements on the constraints on \Asz\ (28\%) and \widthparam\ (23\%).

Because of the consistency of the two calibration datasets, we combine them into a joint mass calibration analysis. We observe that the SZE normalization \Asz\ remains close to the value favored by the $\sigma_v$ measurements, while its 68\% confidence region decreases by roughly 20\% compared to the individual results. This impact on \Asz\ is the best improvement on the SZE parameters we observe when combining the mass calibrators. The constraints on \Om\ and $\sigma_8$ lie between the individual results with similar uncertainties. However, \widthparam\ clearly benefits from the combined mass information, and its uncertainty is 10\% (23\%) smaller than when using the individual \Yx\ (\sigmav) calibration data.


\subsection{\LCDM\ Results with WMAP9}
\label{sec:SPTcl_WMAP9}

\begin{figure}[htb]
\begin{center}
\includegraphics[width=\columnwidth]{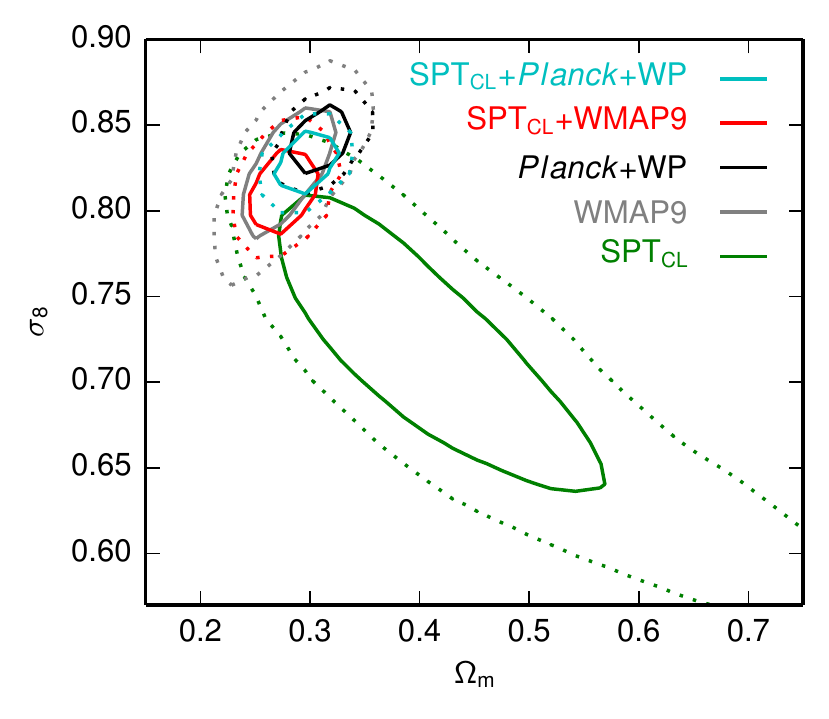}
\caption{Likelihood contours (68\% and 95\%) in \Om-$\sigma_8$ space for SPT clusters with \sigmav\ and \Yx\ (\SPTcl), CMB from WMAP9 and \planck+WP, and the combination of clusters with CMB data. The independent cluster and CMB constraints overlap, and their approximate orthogonality make them particularly complementary. We quantify the agreement between \SPTcl\ and WMAP9 (\planck+WP) to be $1.3\sigma$ ($1.9\sigma$) (see Section~\ref{sec:SPTcl_WMAP9}). Accounting for a single massive neutrino ($m_\nu =0.06$~eV) shifts these values to 1.0$\sigma$ (1.5$\sigma$); treating the sum of neutrino masses as a free parameter yields 0.7$\sigma$ (1.1$\sigma$).}
\label{fig:Omegam-sigma8}
\end{center}
\end{figure}

We now compare the results from our cluster data with constraints from CMB anisotropies as obtained from WMAP9.
The probability distributions of the cluster datasets and WMAP9 overlap, indicating agreement between both sets of constraints (see also Figure~\ref{fig:Omegam-sigma8}).
Moreover, the parameter degeneracies in the \Om-$\sigma_8$ space for clusters are nearly orthogonal to the ones of CMB data.

We quantify the agreement between two datasets by testing the degree to which their probability distributions $P(\vect{x})$ overlap in some parameter space $\vect{x}$.  We measure this by first drawing representative samples of points $\left\{\vect{x_1}\right\}$ and $\left\{\vect{x_2}\right\}$ from the two probability distributions $P_1(\vect{x})$ and $P_2(\vect{x})$.  We then compute the distances between pairs of sampled points $\vect{\delta}\equiv \vect{x_1} - \vect{x_2}$ and estimate the probability distribution $P_\delta$ from this ensemble $\left\{\vect{\delta}\right\}$.  We then evaluate the likelihood $p$ that the origin lies within this distribution:
\begin{equation}
p = \int_{S} d\vect{y}\,  P_\delta(\vect y)
\end{equation}
where the space $S$ is that where $P_\delta<P_\delta(\vect 0)$, and $P_\delta(\vect 0)$ is the probability at the origin.
We convert $p$ to a significance assuming a normal distribution.
Within the PMC fitting procedure used to obtain the probability distributions $P$, each sample point $\vect x$ is assigned a weight. We calculate the agreement between two distributions using the method presented above, assigning each point $\vect{\delta}$ a weight that is the product of the weights of the points $\vect{x_1}$ and $\vect{x_2}$.

We apply this method in the two-dimensional \Om-$\sigma_8$ space. Within our baseline model that assumes massless neutrinos we report good consistency ($1.3 \sigma$) between the results from our cluster sample and from WMAP9. Changing the baseline assumptions to account for one massive neutrino with mass $m_\nu=0.06\,\mathrm{eV}$ decreases the tension to $1.0\sigma$. We note that this increase in neutrino mass shifts CMB constraints towards lower values of $\sigma_8$ by about $\Delta \sigma_8\approx-0.012$ while having negligible impact on the cluster constraints. We fit for the sum of neutrino masses in Section~\ref{sec:massiveneutrinos}; this further reduces the tension.

\begin{deluxetable*}{lcccc}[htb]
\tablecaption{Impact of \sigmav\ and/or \Yx\ mass calibration on results from SPT clusters \Nxiz+WMAP9.}
\tablehead {\colhead{Dataset} & \colhead{\Asz} & \colhead{\Om} & \colhead{$\sigma_8$} & \colhead{\widthparam}}
\startdata
\Nxiz+WMAP9	& $3.59^{+0.60}_{-1.04}$ & $0.284\pm0.027$ & $0.823\pm0.026$ & $0.835\pm0.047$ \\[6pt]
\Nxiz+WMAP9+\sigmav & $3.51^{+0.65}_{-0.63}$ & $0.288\pm0.022$ & $0.824\pm0.020$ & $0.840\pm0.035$ \\[6pt]
\Nxiz+WMAP9+\Yx & $3.85^{+0.62}_{-0.66}$ & $0.273\pm0.019$ & $0.811\pm0.019$ & $0.813\pm0.032$ \\[6pt]
\Nxiz+WMAP9+\Yx+\sigmav & $3.79^{+0.57}_{-0.63}$ & $0.276\pm0.018$ & $0.812\pm0.017$ & $0.817\pm0.027$ \\
\enddata
\tablecomments{These are fully marginalized constraints. The results from \Nxiz+\Yx+\sigmav+WMAP9 are presented in more detail in Table~\ref{tab:LCDM}.}
\label{tab:SZWMAP}
\end{deluxetable*}

Given the overlap between the probability distributions from our clusters and WMAP9 we combine the datasets to break degeneracies and thereby tighten the constraints.  In Table~\ref{tab:SZWMAP}, we show how the combination of the \Nxiz\ cluster sample with WMAP9 data benefits from the additional mass calibration from \sigmav\ and/or \Yx. It is clear that, even if the cosmological constraints are dominated by the CMB data, the mass calibration from either observable leads to tighter constraints on all four parameters shown in the table. We also observe that the constraints on the cosmological parameters \Om, $\sigma_8$, and \widthparam\ obtained when including \Yx\ data are systematically lower by about half a $\sigma$ than results obtained without these data; the constraints on \Asz\ are higher. These shifts correspond to lower cluster masses; we will come back to this in Section~\ref{sec:masses}.

When adding the WMAP9 data to our full cluster sample \SPTcl\ we observe shifts in the SZE scaling relation parameters, as shown in Table~\ref{tab:LCDM}. There is a decrease in the SZE normalization \Asz\ by 19\%, and the uncertainty tightens by 42\%. We further observe a notable shift in the redshift evolution \Csz\ towards a lower value at the $1\sigma$ level. This is due to the degeneracy between \Csz\ and \Om, as the latter also shifts significantly when the WMAP9 data are added. The remaining scaling relation parameters do not benefit from the additional data.  Conversely, the SPT cluster data improve the cosmological constraints from the WMAP9 data by reducing the uncertainty on \Om\ by 36\%, on $\sigma_8$ by 33\%, and on \widthparam\ by 47\%.  Figure~\ref{fig:Omegam-sigma8} shows how the combination of the datasets leads to improved constraints due to the nearly orthogonal parameter degeneracies of the individual results (red contours in figure).

Finally, we add data from BAO and SNIa which carry additional information on cosmic distances. As expected, we see a further tightening of the constraints on $\Omega_\textrm m=0.292\pm0.011$ and $H_0=68.6\pm1.0$~km~s$^{-1}$~Mpc$^{-1}$.


\subsection{\LCDM\ Results with \planck+WP}
In Figure~\ref{fig:Omegam-sigma8}, we also show the constraints in the  \Om-$\sigma_8$ plane from \planck+WP and report a mild $1.9 \sigma$ tension between our cluster sample and this CMB dataset. The tension is slightly larger than when comparing the clusters to WMAP9.
The \planck+WP data favor a larger value of $\sigma_8$ than our cluster sample. Assuming one massive neutrino with mass $m_\nu=0.06\,\mathrm{eV}$ relaxes the tension to $1.5\sigma$.

We proceed and combine our cluster sample with the CMB data from \planck+WP. This data combination prefers a value for $\sigma_8$ that is about $1\sigma$ lower than suggested by the CMB data. Adding our cluster sample to \planck+WP leads to improvements on the constraints on \Om, $\sigma_8$ and \widthparam, all on the order of 15\% (see Table~\ref{tab:LCDM}, and black/cyan contours in Figure~\ref{fig:Omegam-sigma8}).

We add BAO and SNIa data to further improve the cosmological constraints, and measure $\Omega_m=0.297\pm0.009$, $\sigma_8=0.829\pm0.011$, \widthparam$=0.855\pm0.016$, and $H_0=68.3\pm0.8$~km~s$^{-1}$~Mpc$^{-1}$.  These represent improvements of 18\% (\Om), 8\% ($\sigma_8$), and 11\% (\widthparam) over the constraints from \planck+WP+BAO+SNIa without \SPTcl.  In addition, these represent improvements of 18\% (\Om), 31\% ($\sigma_8$), 20\% (\widthparam) and 20\% ($H_0$) over the corresponding parameter uncertainties when using WMAP9 instead of \planck+WP.


\begin{figure}[htbp]
\begin{center}
\includegraphics[width=\columnwidth]{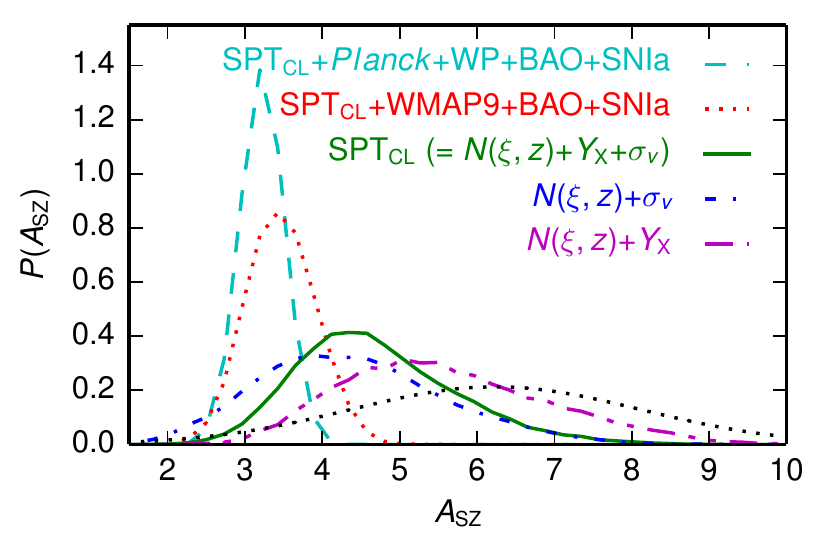}
\caption{Posterior probability distributions for the normalization \Asz\ of the SZE $\zeta$-mass relation for different combinations of mass calibration, CMB, and additional datasets. The Gaussian prior is shown by the black dashed curve. Note the systematic trend towards lower \Asz\ values and smaller uncertainty when adding external cosmological data, corresponding to an increase in the characteristic scale of SPT cluster masses by $\sim$44\% from \Nxiz+\Yx\ (magenta) to \SPTcl+CMB+BAO+SNIa (cyan/red).}
\label{fig:Asz}
\end{center}
\end{figure}


\subsection{Impact on Cluster Masses}
\label{sec:masses}
Combining the mass calibration from \Yx\ with \sigmav\ data and further with CMB data leads to shifts in the SZE scaling relation parameters which ultimately shift the mass estimates of the clusters. As shown in Figure~\ref{fig:Asz}, there is a systematic increase of the cluster mass scale as we move from X-ray to dispersion only calibration, further on to \Yx+\sigmav\ and finally on to analyses of our \SPTcl\ dataset in combination with external datasets (remember that a decrease in \Asz\ corresponds to an increase in cluster mass, see Equation~\ref{eq:scalrelSZ}). Also, it is clear that the constraints on the SZE normalization \Asz\ obtained when including CMB data are much stronger than the constraints from the cluster data alone. The Gaussian prior on \Asz\  is in some tension with the \Asz\ constraints after including the CMB data. In this case, we note that the recovered values of \Asz\ do not significantly change when removing the prior, because it is much broader than the recovered constraints.

We quantify the agreement between these distributions in the space of \Asz\ in a way equivalent to the one presented in Section~\ref{sec:SPTcl_WMAP9}. We find that the results from both \Yx\ and \sigmav\ mass calibration are consistent at the $0.6\sigma$ level.  There is a mild tension ($1.9\sigma$) between mass calibration from \Yx\ and \SPTcl+\planck+WP+BAO+SNIa, while the mass calibration from \sigmav\ is consistent with the multi-probe dataset at the $0.8\sigma$ level. These shifts would approximately correspond to an increase in the preferred cluster mass scale by 44\% and 23\%, respectively, when using the multi-probe dataset. Note that there are shifts in \Bsz\ and \Csz\ when adding CMB data to our cluster sample which add a slight $\xi$ (or equivalently mass), and redshift dependence to this comparison of cluster masses.

On average, our cluster mass estimates are higher by 32\% than our previous results in \citet{reichardt13}, primarily driven by using new CMB and BAO datasets. Relative to \citet{reichardt13}, we have updated the CMB data set from WMAP7 and SPT \citep{komatsu11, keisler11} to \planck+WP \citep{planck13-25,planck13-26}, and also updated the BAO dataset from \citet{percival10} to a combination of three measurements \citep{beutler11, anderson12,padmanabhan12}. The new datasets have led to more precise constraints on the cosmological parameters, in particular \widthparam, and drive shifts in the preferred cluster mass scale through \Asz, to improve consistency between the cluster data set and the cosmological constraints. For example, using WMAP9 data instead of \planck+WP+BAO+SNIa leads to an average 11\% decrease of the cluster masses. Finally, we observe an increase in the slope \Bsz\ as compared to \cite{reichardt13} which reduces the mass change to only $\sim15\%$ on the high-mass end of the sample.


\subsection{Goodness of Fit of Cluster Data}
\label{sec:goodnessoffit}

Our analysis to this point has focused on extracting parameter confidence regions that emerge from different combinations of our cluster sample with external datasets.  We observed shifts especially in the SZE scaling relation parameters when switching among the different data combinations. In the following, we investigate whether the adopted SZE mass-observable scaling relation parametrization is adequate for describing the cluster sample.  We execute two tests: (1) we evaluate the goodness of fit of the SZE selected clusters in the $\xi$-$z$ plane, and (2) we compare the predicted values for the follow-up observables \Yx\ and $\sigma_v$ to their actual measurements. Both tests are performed adopting parameter values at the best-fit location in cosmological and scaling relation parameter space from the \SPTcl+\planck+WP+BAO+SNIa analysis.

\begin{figure}[htb]
\begin{center}
\includegraphics[width=\columnwidth]{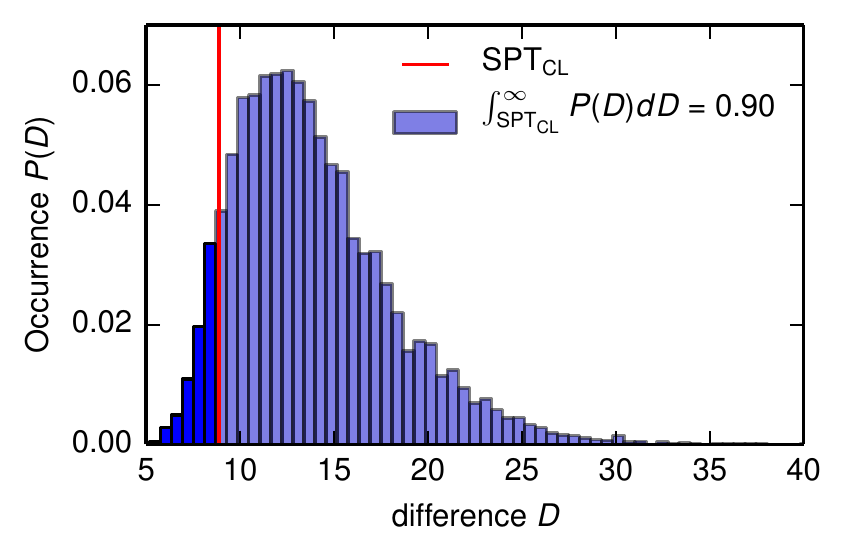}
\caption{The goodness of fit of our cluster dataset to the best-fit cosmological model is evaluated using a two-dimensional KS test on the distribution of clusters in SZE signature $\xi$ and redshift $z$ (see Section~\ref{sec:goodnessoffit}).  The blue histogram is the expected distribution of differences $D$ between the observations and the model for an ensemble of 10,000 simulated realizations of the best-fit cosmology. The \SPTcl\ dataset is marked by the red line and exhibits no tension with the parametrization from the best-fit model.}
\label{fig:2DKS}
\end{center}
\end{figure}

We compare the distribution of the SZE clusters in the observable $\xi$-$z$ plane with its prediction. This is done using a two-dimensional Kolmogorov-Smirnov (KS) test as described in \cite{press92}: At the location of each cluster in $\xi$ and $z$ space, we split the observational space into four quadrants, and calculate the absolute difference between the number of clusters and the number predicted by the model within that area. The largest of these $4\times N_\textrm{cl}$ values is taken as the maximum difference $D$ between the data and the model. We characterize this difference measure by calculating it for 10,000 independent catalogs that we produce using the best-fit cosmology and scaling relation parameters. Figure~\ref{fig:2DKS} contains a histogram of the distribution of differences $D$ from the set of catalogs, and the red line marks the difference for the real sample. This test indicates that there is a 90\% chance of obtaining a larger difference $D$ than observed in our real dataset. We conclude that there is no tension between our SPT cluster sample and the way we model it through the SZE scaling relation parametrization.

\begin{figure}[htb]
\begin{center}
\includegraphics[width=\columnwidth]{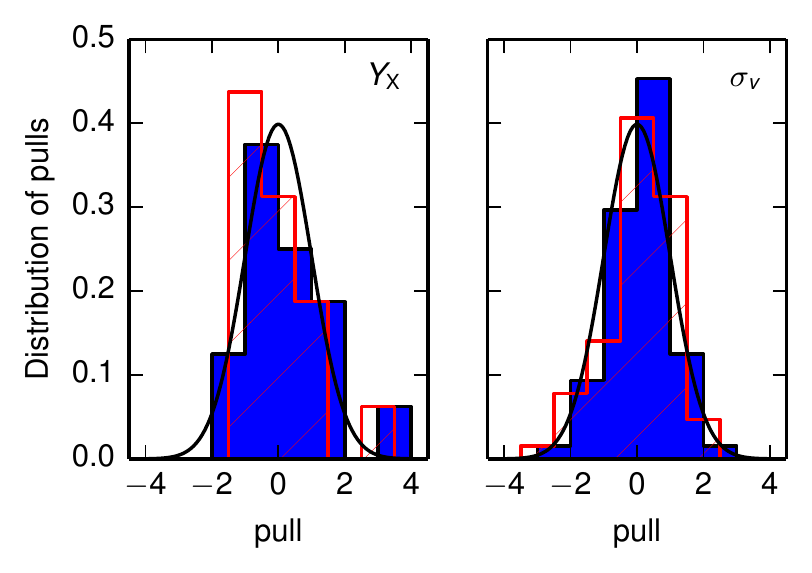}
\caption{Difference of the X-ray and dispersion follow-up mass measurements and their predictions from SZE. We show the distribution of pulls (see Section~\ref{sec:goodnessoffit}), and the expected Gaussian distribution in black. The result obtained from clusters alone is shown in blue, and the combined results from all cosmological probes are shown in red.  A KS test indicates there is no tension between our cluster mass calibration data and the expected mass distribution in the best-fit cosmology.}
\label{fig:1DKS}
\end{center}
\end{figure}

We now go one step further and ask whether there is tension between the predicted values for the follow-up observables \Yx\ and $\sigma_v$ and their actual measurements. Remember that the predicted probability distributions are obtained from the observed SZE signal $\xi$ according to Equation~\ref{eq:P_O_xi}. For each cluster, we calculate the percentile of the observed value in its predicted distribution. We get a distribution of percentiles which we convert to a distribution of pulls \citep{eadie83,lyons89} using the inverse error function:
\begin{equation}
\mathrm{pull} = \sqrt{2}\times \mathrm{erf}^{-1}(2\times\mathrm{percentile}-1).
\end{equation}
This distribution is finally compared to a normal distribution of unit width centered at zero using the KS test. In Figure~\ref{fig:1DKS} we show the distribution of pulls for the \Yx\ and \sigmav\ measurements. For each observable, we show the distribution for two different sets of cosmological and scaling relation parameters: (1) the results obtained from clusters with mass calibration only, and (2) the results from clusters with mass calibration combined with the external cosmological probes. In all 4 cases, the KS test provides $p$-values in the range $0.1<p<0.8$, indicating no tension between the predicted follow-up mass observables and their measurements.
This is an interesting observation given the shifts we observe in the scaling relation and cosmological parameters when adding CMB data to the cluster sample. It shows that the adopted form of the SZE mass-observable scaling relation has enough freedom to compensate for the shifts in cosmological parameters.  With a larger cluster and mass calibration dataset we could expect to make a more precise consistency test of the data and our adopted scaling relation parametrization.

\subsection{Dark Energy Equation of State}

The first extension of the \LCDM\ model we analyze is the flat $w$CDM cosmology which includes the Dark Energy equation of state parameter $w$.  As the Dark Energy becomes relevant only in the late Universe and affects the cluster mass function through its impact on the cosmological growth rate and volume we expect our cluster sample to provide an important contribution in constraining its nature.

Analyzing our cluster sample using priors on $H_0$ and BBN, we obtain $w=-1.5\pm0.5$. This measurement is compatible with external constraints from WMAP9+$H_0$ ($w=-1.13\pm0.11$) and \planck+WP+BAO \citep[$w=-1.13\pm0.25$, 95\% confidence limits;][]{planck13-26}, and consistent with the \LCDM\ value $w=-1$. 
Remember that the results obtained from clusters might in principle be subject to systematics in the mass estimates, while, on the other hand, the CMB anisotropy measurements are most sensitive to the characteristics of the Universe at $z\sim1100$, and the distance measurements are subject to their own systematics.

Combining datasets breaks degeneracies and leads to tighter constraints.
When adding our \SPTcl\ sample to the WMAP9+$H_0$ data, we measure $w=-1.07\pm0.09$, or an 18\% improvement over the constraint without clusters.  Combining our cluster sample with \planck+WP+BAO+SNIa ($w=-1.051\pm0.072$) data leads to an even tighter constraint, and we measure $w=-0.995\pm0.063$ (12\% improvement, see also Table~\ref{tab:LCDMext}).

\begin{deluxetable*}{lccccc}[htb]
\tablecaption{Constraints on extensions of flat \LCDM\ cosmology from the \\\SPTcl+\planck+WP+BAO+SNIa data combination.}
\tablehead {\colhead{Parameter} & \colhead{$w$CDM} & \colhead{$\nu$CDM} & \colhead{$\gamma$+\LCDM} & \colhead{$\gamma$+$\nu$CDM} & \colhead{$\gamma$+$w$CDM}}
\startdata
\Om\			& $0.301\pm0.014$ 	& $0.309\pm0.011$ & $0.302\pm0.010$	& $0.309\pm0.012$ & $0.301\pm0.014$\\[6pt]
$\sigma_8$	& $0.827\pm0.024$ 	& $0.799\pm0.021$ & $0.793^{+0.046}_{-0.075}$ & $0.796^{+0.057}_{-0.080}$ & $0.794^{+0.054}_{-0.078}$ \\[6pt]
$H_0$~(km~s$^{-1}$~Mpc$^{-1}$)		& $68.1\pm1.6$	& $67.5\pm0.9$ & $68.2\pm0.8$	& $67.5\pm0.9$ & $68.3\pm1.6$ \\[6pt]
$w$			& $-0.995\pm0.063$  	& $(-1)$ & $(-1)$			& $(-1)$ & $-1.007\pm0.065$ \\[6pt]
\summnu~(eV)	& $(0)$			& $0.148\pm0.081$ & $(0)$	& $0.143^{+0.066}_{-0.100}$ & $(0)$ \\[6pt]
\summnu~(eV), 95\% CL	& $(0)$			& $<0.270$ & $(0)$	& $<0.277$ & $(0)$ \\[6pt]
$\gamma$	& $(0.55)$			& $(0.55)$ & $0.72\pm0.24$	& $0.63\pm0.25$ & $0.73\pm0.28$
\enddata
\tablecomments{These are fully marginalized constraints.}.
\label{tab:LCDMext}
\end{deluxetable*}


\subsection{Massive Neutrinos}
\label{sec:massiveneutrinos}

We now extend the \LCDM\ model and include the sum of neutrino masses \summnu\ as a free parameter. We will refer to this model as $\nu$CDM in the following, and we assume three degenerate mass neutrino species.

Massive neutrinos are still relativistic at the epoch of recombination and hence do not significantly affect the structure of CMB anisotropies \citep[as long as $m_\nu<0.6$~eV for each species, ][]{komatsu09}. In the late Universe, massive neutrinos contribute to \Om\ but do not cluster in structures smaller than their free streaming length, leading to a lower $\sigma_8$. Therefore, results from CMB anisotropy data exhibit a strong degeneracy between \summnu\ and $\sigma_8$. Using the \planck+WP+BAO+SNIa data combination we measure $\sum m_\nu = 0.092\pm0.058$~eV and an upper limit $\sum m_\nu <0.182$~eV (95\% confidence limit, hereafter CL).

Galaxy clusters are ideal probes for measuring $\sigma_8$ and therefore represent a valuable piece of information when constraining the $\nu$CDM model. When adding our \SPTcl\ sample to the dataset, we observe that the mean of the recovered \summnu\ increases significantly; we measure $\sum m_\nu = 0.148\pm0.081$~eV, and an upper limit $\sum m_\nu < 0.270$~eV (95\% CL). As discussed earlier, our cluster sample prefers lower values for $\sigma_8$ than the CMB data, which here leads to increased neutrino masses due to their degeneracy with $\sigma_8$.
The results on $\nu$CDM from the full data combination are also shown in Table~\ref{tab:LCDMext}.

We recalculate the difference between results from \SPTcl\ and CMB data as in Section~\ref{sec:SPTcl_WMAP9}, but we now adopt our best-fit sum of neutrino masses $\sum m_\nu = 0.148$~eV. This decreases the tension to 0.7$\sigma$ for WMAP9, and 1.1$\sigma$  for \planck+WP.


\subsection{Testing the Cosmological Growth of Structure}
\label{sec:structuregrowth}
Our constraints on the Dark Energy equation of state parameter confirm once more that the flat \LCDM\ model provides an excellent fit to the best currently available cosmological data. However, it still remains unclear what exactly is causing the accelerating expansion in the present epoch. Possible explanations include a new energy component or a modification of gravity on large scales. While measurements of CMB anisotropies and cosmic distances (BAO and SNIa) have proven extremely useful for probing the expansion history of the Universe, galaxy clusters provide a unique probe for testing its growth history. Combining these tests allows for an interesting consistency test of General Relativity (GR) on large scales \cite[e.g.,][]{rapetti13}.

\subsubsection{Parametrized Growth of Structure}
We parametrize the linear growth rate of density perturbations $f(a)$ at late times as a power law of the matter density \cite[e.g.,][]{peebles80,wang98}
\begin{equation}
f(a) \equiv \frac{d \ln \delta}{d\ln a} = \Omega_m(a)^\gamma
\end{equation}
where $\gamma$ is the cosmic growth index and $\delta\equiv \delta\rho_m/\langle\rho_m\rangle$ is the ratio of the comoving matter density fluctuations and the mean matter density. Solving for $\gamma$ and assuming GR one obtains
\begin{equation}
\gamma_\textrm{GR} \approx \frac{6-3(1+w)}{11-6(1+w)}
\end{equation}
where the leading correction depends on the dark energy equation of state parameter $w$ and so $\gamma_\textrm{GR}=0.55$ for a cosmological constant with $w=-1$.
Normalizing the parametrized cosmic growth factor $D(z) \propto \delta(z)$ at some high redshift $z_\textrm{ini}$ 
we can express it as
\begin{equation} \label{eq:Dini}
D_\textrm{ini}(z) = \frac{\delta(z)}{\delta(z_\textrm{ini})} =  \delta(z_\textrm{ini}) ^{-1}  \exp  \int d\ln a \; \Omega_\textrm{m}(a)^\gamma
\end{equation}
and the parametrized matter power spectrum becomes
\begin{equation}
\mathrm P(k,z) = P(k,z_\textrm{ini}) D_\textrm{ini}^2(z).
\label{eq:parametrizedPS}
\end{equation}
Note that the complete wavenumber-dependence is contained in $P(k,z_\textrm{ini})$ while the growth factor $D_\textrm{ini}(z)$, which now depends on $\gamma$, evolves with redshift only.

In our analysis, we choose an initial redshift of $z_\textrm{ini}=10$ as a starting point for the parametrized growth which corresponds to an era well within matter domination when $f(a)=1$ is a very good approximation.
We modify the likelihood code presented in Section~\ref{sec:clustermassfunction} so that the matter power spectrum at redshift $z_\textrm{ini}$ is provided by CAMB and then evolves depending on the growth index $\gamma$ according to Equations~\ref{eq:Dini} and~\ref{eq:parametrizedPS}.

We note that this parametrization is in principle degenerate with a cosmological model containing neutrino mass as a free parameter; given a particular power spectrum constrained by the CMB anisotropies at very high redshift, variations in both neutrino mass and $\gamma$ modify the low-redshift power spectrum. However, the SPT sample spans a broad redshift range which should ultimately allow one to differentiate between the two effects.

\subsubsection{Constraints on the Cosmic Growth Index}

\begin{figure}[htbp]
\begin{center}
\includegraphics[width=\columnwidth]{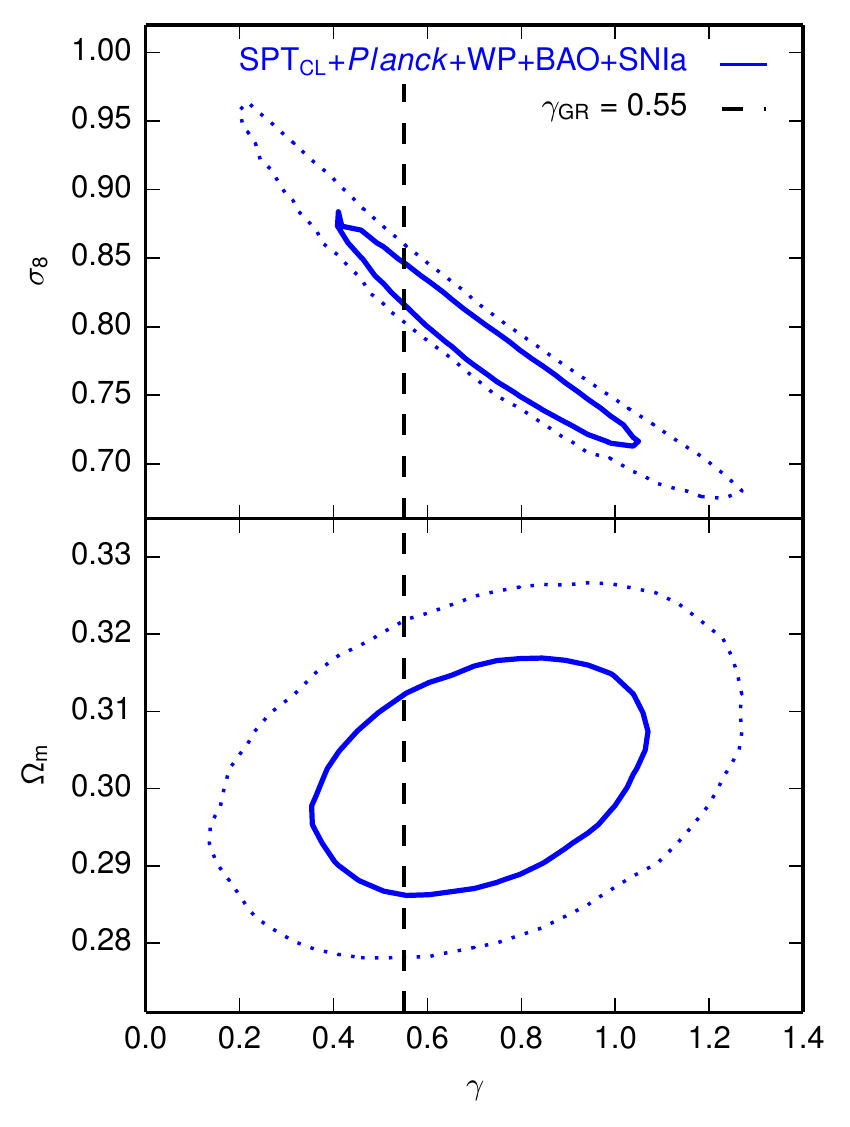}
\caption{$\gamma$+\LCDM: Likelihood contours (68\% and 95\%) for the growth index $\gamma$ and $\sigma_8$ (top), and $\gamma$ and \Om\ (bottom). The prediction by GR $\gamma_\textrm{GR}=0.55$ is indicated by the dashed line. The strong degeneracy between $\gamma$ and $\sigma_8$ is clear. We measure $\gamma=0.72\pm0.24$, indicating no tension with the growth rate predicted by GR.}
\label{fig:gamma}
\end{center}
\end{figure}

We fit for a spatially flat \LCDM\ model with the additional degree of freedom $\gamma$ (we will refer to this model as $\gamma$+\LCDM).
Using our \SPTcl\ sample with BBN and $H_0$ priors, we get results that are consistent with the prediction of GR, $\gamma_\textrm{GR}=0.55$. However, the uncertainty on $\gamma$ is large, and the 68\% confidence interval is $[-0.2, 0.7]$.
We tighten the constraints by including the CMB dataset which serves as a high-redshift ``anchor'' of cosmic evolution.  To isolate the constraining power clusters have on growth of structure, we choose not to use the constraints on $\gamma$ that come from the Integrated Sachs-Wolfe (ISW) effect, which has an impact on the low $l$ CMB temperature anisotropy. Regardless, we would expect the additional constraints on $\gamma$ from the ISW to be less constraining than the cluster-based constraints presented here \citep[see, e.g., ][]{rapetti10}.
We further use distance information from BAO and SNIa.  As presented in Table~\ref{tab:LCDMext}, we find $\gamma = 0.72\pm0.24$, which agrees with the prediction of GR. In Figure~\ref{fig:gamma}, we show the two-dimensional likelihood contours for $\gamma$ and the most relevant cosmological parameters \Om\ and $\sigma_8$. The degeneracy between $\gamma$ and \Om\ is weak. We see a strong degeneracy with $\sigma_8$, as would be expected given the dependence of $\sigma_8$ on growth history.

Our constraints are weaker than those obtained from an X-ray cluster sample \citep{rapetti13}. Using 238 clusters from different X-ray catalogs together with CMB anisotropy data from the 5-year WMAP release these authors obtain $\gamma=0.415\pm0.127$.

We also consider a $\gamma$+$\nu$CDM cosmological model, where we additionally allow a non-zero sum of the neutrino masses.
There is only a mild degeneracy between $\gamma$ and \summnu, which does not significantly degrade our constraints on cosmic growth or neutrino masses (see upper panel of Figure~\ref{fig:gamma} and Table~\ref{tab:LCDMext}). However, the best-fit value for $\gamma$ shifts by $\sim 0.5\sigma$ closer to the GR value.

\begin{figure}[htbp]
\begin{center}
\includegraphics[width=\columnwidth]{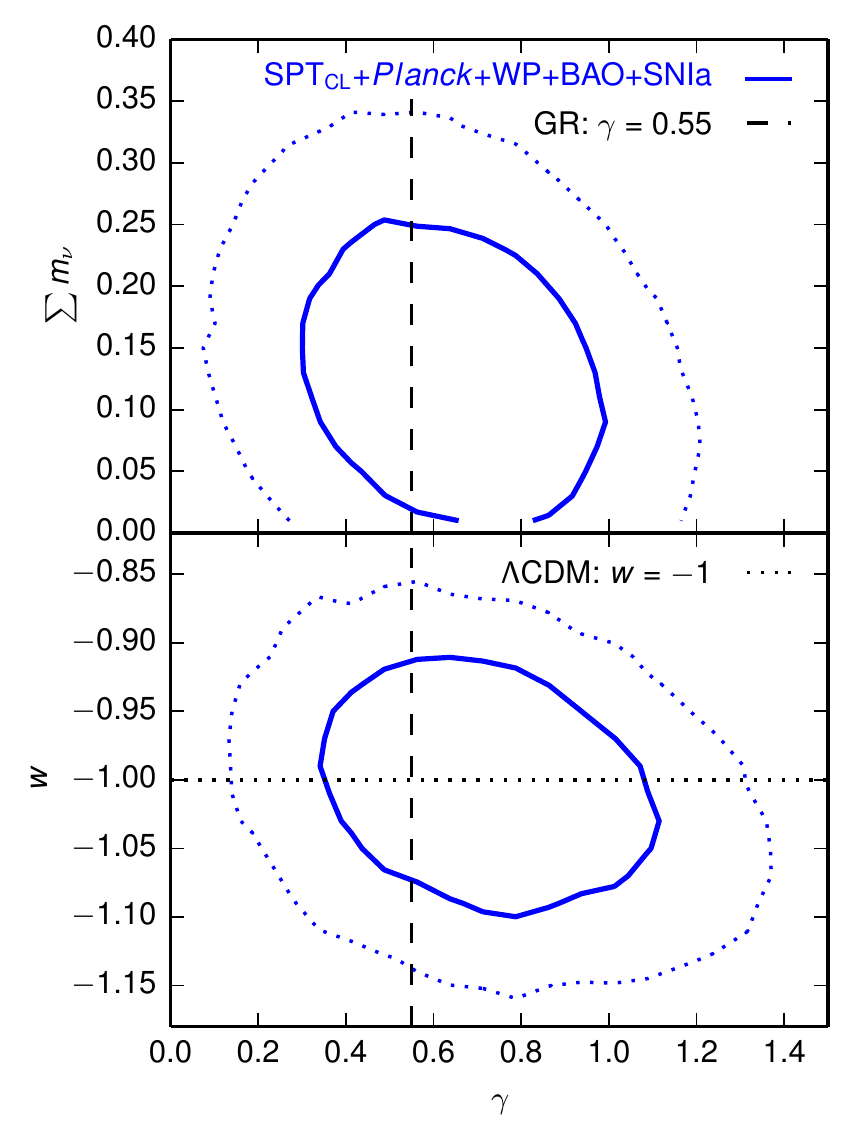}
\caption{Likelihood contours (68\% and 95\%) for $\gamma$+\LCDM\ with additional one-parameter extensions \summnu\ (top), and $w$ (bottom). The prediction for $\gamma$ by GR and the \LCDM\ value for $w$ are indicated by the lines. The cosmological datasets combined exhibit no tension with a GR$+\Lambda$CDM description of the Universe.}
\label{fig:wgamma}
\end{center}
\end{figure}

Finally, we consider a $\gamma$+$w$CDM cosmological model, where we fix $\sum m_\nu=0$~eV, and allow a varying Dark Energy equation of state parameter $w$. In doing so we can simultaneously account for possible departures from the standard cosmic growth history as well as departures from the expansion history as described by the \LCDM\ model.  As presented in Table~\ref{tab:LCDMext}, the results show consistency with the fiducial values $\gamma_\textrm{GR}=0.55$ and $w_{\Lambda\textrm{CDM}}=-1$. Joint parameter constraints are shown in the bottom panel of Figure~\ref{fig:wgamma}.  This combined test confirms that the standard cosmological model accurately describes the evolution of the cosmic expansion and structure formation throughout a wide redshift and distance range.

\section{Summary}
\label{sec:conclusions}

We use an SZE selected galaxy cluster sample from 720~deg$^2$ of the SPT-SZ survey in combination with follow-up data from optical spectroscopy and X-ray observations to carry out a calibration of the SPT mass-observable relation.  This work improves on previous analyses by the inclusion of the velocity dispersion data.

We present a method to fit for the SPT mass-observable relation through comparison of the SZE observable to the external calibrators $\sigma_v$ and/or \Yx. The method accounts for selection effects in the SPT cluster survey, for intrinsic scatter in the mass-observable scaling relations, for observational uncertainties, and for uncertainties in the scaling relation parameters. With this method we compute the likelihood for the cluster counts in the space of $\xi$ and $z$, and for the mass calibration using measurements in the follow-up observables.

Before combining the \Yx\ and \sigmav\ mass calibration datasets we show that their individual constraints on the SPT $\zeta$-mass scaling relation parameters are comparable, agreeing at the $0.6\sigma$ level. Given the different nature of \Yx\ and $\sigma_v$ and their different calibration schemes, we argue that this agreement is a useful crosscheck of systematics present in either calibrating dataset.  Combining the mass calibration datasets leads to an improvement of the constraints on \Asz\ and \widthparam.  Cosmological constraints from SPT clusters with external BBN and $H_0$ priors differ from the independent CMB anisotropy constraints from WMAP9 (\planck+WP) at the $1.3\sigma$ ($1.9\sigma$) level (see Figure~\ref{fig:Omegam-sigma8} and Table~\ref{tab:LCDM}).  Accounting for the impact of one massive neutrino ($m_\nu=0.06$~eV) reduced the differences to 1.0$\sigma$ (1.5$\sigma$). 

Combining our SPT cluster sample with CMB data from WMAP9, we show that the mass calibration from \sigmav\ or \Yx\ lead to tighter constraints on key cosmological parameters; the use of both mass calibration datasets together furthers tightens these constraints.  Throughout the different combinations of cluster mass calibration and external data, we observe that the cluster mass scale from dispersions is higher than the one inferred from \Yx. As we summarize in Figure~\ref{fig:Asz}, the SZE scaling relation normalization \Asz\ obtained using the multi-probe dataset is in better agreement with the \sigmav\ calibration results (0.8$\sigma$) than with the \Yx\ calibration results (1.9$\sigma$).
Analyzing the cluster sample with data from \planck+WP, BAO, and SNIa, we find that the average cluster masses in this work have increased by $\sim$32\% relative to \cite{reichardt13}, primarily driven by the use of new CMB and BAO datasets, which prefer a \LCDM\ cosmology with a higher \widthparam.

Assuming a flat \LCDM\ model, and using the SPT cluster catalog, \sigmav\ and \Yx\ mass calibration, and external data from \planck+WP, BAO, and SNIa, we measure $\Omega_\textrm m=0.299\pm0.009$, $\sigma_8=0.829\pm0.011$, and $\sigma_8\left(\Omega_\textrm m/0.27\right)^{0.3}=0.855\pm0.016$.  These correspond to 18\% (\Om), 8\% ($\sigma_8$), and 11\% (\widthparam) improvements over the constraints from \planck+WP+BAO+SNIa without \SPTcl.

We execute two goodness of fit tests to evaluate whether the adopted SZE mass-observable scaling relation parametrization is adequate to describe our cluster sample.  As shown in Figure~\ref{fig:2DKS}, there is good agreement between the distribution of the observed cluster sample in $\xi$ and $z$, and the prediction by the model.
We also find good agreement between the predicted SZE mass estimates, and the follow-up mass measurements, using either \sigmav\ and \Yx\ (see Figure~\ref{fig:1DKS}).

We examine an extension of the standard \LCDM\ model by adding the Dark Energy equation of state parameter $w$.  Our results are all compatible with $w=-1$, and our best constraint is $w=-0.995\pm0.063$, which we obtained from our cluster sample in combination with \planck+WP, BAO, and SNIa (12\% improvement after adding \SPTcl). We consider another extension to \LCDM\ in which we fit for the sum of neutrino masses, and find  $\sum m_\nu = 0.148\pm0.081$~eV, with $\sum m_\nu < 0.270$~eV (95\% CL).

We then allow for another additional cosmological degree of freedom by parametrizing the cosmic growth rate. The growth index is constrained to $\gamma = 0.72\pm0.24$ when assuming a \LCDM\ background.  This agrees with the GR prediction $\gamma_\textrm{GR}=0.55$, indicating that the growth of structure is correctly described by GR.
We consider the effect on $\gamma$ when additionally allowing a non-zero sum of the neutrino masses, and find only a weak degeneracy between the two parameters, with relatively small changes in the constraints on $\gamma$ and \summnu.
Finally, we consider a $\gamma$+$w$CDM model, and allow both $\gamma$ and $w$ to vary. We recover results ($\gamma=0.73\pm0.28$ and $w=-1.007\pm0.065$) that are consistent with the predictions of the standard GR+\LCDM\ cosmological model.

Velocity dispersions haven proven to be useful follow-up mass calibrators in our analysis. However, much of their constraining power relies on a precise knowledge of the scaling relation normalization \Adisp, which we assume to be calibrated to within 5\% from $N$-body simulations \citep{saro13}. When relaxing this prior to 10\% in an analysis that uses only the SZE clusters and the measured \sigmav's, the constraint on the SZE normalization \Asz\ degrades by 25\%, and the cosmological constraints relax modestly (14\% on  \widthparam). A better knowledge of the systematics in the \sigmav\ mass-observable relation, in particular the galaxy velocity bias, is therefore crucial for obtaining better constraints from ongoing and future galaxy cluster surveys. This improved knowledge could be obtained with detailed numerical simulations as well as large spectroscopic datasets.

The next steps in the SPT mass calibration consist of the inclusion of weak lensing masses and a larger number of dispersions from an ongoing program on Gemini focused at $z<0.8$ and a complementary program focused at $z>0.8$ on the VLT.  In addition, X-ray observations of a sample of approximately $\sim100$ systems with \chandra\ and \xmm\ are complete. Improved calibration of the mass-observable relations for \Yx\ and $\sigma_v$ would lead to stronger cosmological constraints. Combined analyses of these calibration data together with the full SPT cluster sample \citep{bleem14} will enable significant progress in cluster studies of cosmology and structure formation.

\acknowledgments

We acknowledge the support of the DFG Cluster of Excellence ``Origin and Structure of the Universe'' and the Transregio program TR33 ``The Dark Universe''. The calculations have been carried out on the computing facilities of the Computational Center for Particle and Astrophysics (C2PAP) and of the Leibniz Supercomputer Center (LRZ).
Optical spectroscopic data from VLT programs 086.A-0741 and 286.A-5021 and Gemini program GS-2009B-Q-16 were included in this work. Additional data were obtained with the 6.5~m Magellan Telescopes, which is located at the Las Campanas Observatory in Chile.
This work is based in part on observations made with the \spitzer\ \textit{Space Telescope}, which is operated by the Jet Propulsion Laboratory, California Institute of Technology under a contract with NASA.
The South Pole Telescope is supported by the National Science Foundation through grant PLR-1248097.  Partial support is also provided by the NSF Physics Frontier Center grant PHY-1125897 to the Kavli Institute of Cosmological Physics at the University of Chicago, the Kavli Foundation and the Gordon and Betty Moore Foundation grant GBMF 947.
Galaxy cluster research at Harvard is supported by NSF grant AST-1009012, and research at SAO is supported in part by NSF grants AST-1009649 and MRI-0723073.
Work at Argonne National Lab is supported by UChicago Argonne, LLC, Operator of Argonne National Laboratory (``Argonne''). Argonne, a U.S. Department of Energy Office of Science Laboratory, is operated under Contract No. DE-AC02-06CH11357.
The McGill group acknowledges funding from the National Sciences and Engineering Research Council of Canada, Canada Research Chairs Program, and the Canadian Institute for Advanced Research.

Facilities: \facility{Gemini-S (GMOS)}, \facility{Magellan: Baade (IMACS)}, \facility{South Pole Telescope}, \facility{\spitzer/IRAC}, \facility{VLT: Antu (FORS2)}

\bibliography{spt}

\appendix

\section{Analysis method and Likelihood function}

We show that the analysis method we use in the present work is equivalent to the method used in previous SPT analyses. Specifically, we show how we separate the mass calibration from the cluster number counts. As presented in Equation~4 in  \cite{benson13}, the expected number density in terms of $\xi$, $z$ and the follow-up observable \Yx\ is 
\begin{equation}
\label{eq:dNdxiBenson}
\frac{dN(\xi,Y_\textrm{X},z |\vect{p})}{d\xi dY_\textrm{X} dz}d\xi dY_\textrm{X} dz = \int dM P(\xi,Y_\textrm{X}|M,z,\vect{p})P(M,z|\vect{p}) \Theta(\xi-5,z-0.3),
\end{equation}
and the likelihood function is evaluated according to Poisson statistics
\begin{equation}
\label{eq:likelihood}
\ln \mathcal L(\vect{p}) = \sum_i \ln \frac{dN(\xi_i,Y_{\textrm{X}i},z_i |\vect{p})}{d\xi dY_\textrm{X} dz} - \int\frac{dN(\xi,Y_\textrm{X},z |\vect{p})}{d\xi dY_\textrm{X} dz}d\xi dY_\textrm{X} dz,
\end{equation}
up to a constant offset, and where the sum over $i$ runs over all clusters in the sample.

We assume no correlated scatter in the different observables, i.e. we assume that $P(\xi,Y_\textrm{X}|M,z,\vect{p})=P(\xi|M,z,\vect{p})P(Y_\textrm{X}|M,z,\vect{p})$ holds, and transform Equation~\ref{eq:dNdxiBenson} into two separate factors; this is the analysis method we use here.
In the following, and for ease of reading, we omit $z$ and $\vect{p}$ (e.g., $P(M) \equiv P(M|z,\vect{p})$), and the selection function $\Theta(\xi-5)$ as it does not depend on mass for a given cluster with measured $\xi$. We use Bayes' theorem twice, e.g. $P(\xi|M)P(M)=P(M|\xi)P(\xi)$.
\begin{eqnarray}
\frac{dN(\xi,Y_\textrm{X},z |\vect{p})}{d\xi dY_\textrm{X} dz} &=& \int dM P(\xi,Y_\textrm{X}|M)P(M) \nonumber\\
	&=& \int dM P(Y_\textrm{X}|M) P(\xi|M)P(M)  \int dM' P(M'|\xi) \nonumber\\
	&=&  \iint dM dM' P(Y_\textrm{X}|M) P(M|\xi)P(\xi)  \frac{P(\xi|M')P(M')}{P(\xi)} \nonumber\\
	&=& \int dM  P(Y_\textrm{X}|M) P(M|\xi)  \int dM' P(\xi|M')P(M') \nonumber\\
	&\equiv& P(Y_\textrm{X} | \xi,z,\vect{p})  \frac{dN(\xi,z |\vect{p})}{d\xi dz}
\end{eqnarray}
With this, the likelihood function we use in this work is
\begin{equation}
\ln\mathcal L(\vect{p}) = \sum_j \ln P(Y_{\textrm{X}j}|\xi_j, z_j,\vect{p}) + \sum_k \ln  \frac{dN(\xi_k,z_k|\vect{p})}{d\xi dz} - \int\frac{dN(\xi,z|\vect{p})}{d\xi dz}d\xi dz
\end{equation}
where the sum over $k$ runs over the full SPT-SZ cluster catalog, and $j$ runs over all clusters with \Yx\ measurements, thereby marginalizing over \Yx\ for clusters without X-ray data. Note that the total number of expected clusters $\int\frac{dN(\xi,z|\vect{p})}{d\xi dz}d\xi dz$ does not depend on \Yx.
The generalization to include the \sigmav\ observable is straightforward.

\end{document}